\documentclass[pdflatex,sn-mathphys-num]{sn-jnl}

\usepackage{graphicx}%
\usepackage{multirow}%
\usepackage{amsmath,amssymb,amsfonts}%
\usepackage{amsthm}%
\usepackage{mathrsfs}%
\usepackage[title]{appendix}%
\usepackage{xcolor}%
\usepackage{textcomp}%
\usepackage{manyfoot}%
\usepackage{booktabs}%
\usepackage{algorithm}%
\usepackage{algorithmicx}%
\usepackage{algpseudocode}%
\usepackage{listings}%
\usepackage{tcolorbox}
\usepackage{paralist}
\usepackage{svg}
\usepackage{ulem}
\usepackage{pifont}
\newcommand{\yes}{\ding{51}}
\newcommand{\no}{\ding{55}}

\usepackage{subcaption}
\usepackage{tcolorbox}           
\usepackage{listings}
\tcbuselibrary{skins, breakable}

\renewcommand{\thefigure}{S\arabic{figure}}
\renewcommand{\thetable}{S\arabic{table}}

\newtcolorbox{agentbox}[1]{
    enhanced,
    breakable,
    colback=white,
    colframe=black!30,
    boxrule=2.0pt,
    sharp corners,
    title={#1},
    fonttitle=\bfseries,
    titlerule=5pt,
    left=0.8em,
    right=0.8em,
    top=0.5em,
    bottom=0.5em,
    boxsep=0pt,
    before skip=1em,
    after skip=1em,
}

\theoremstyle{thmstyleone}%
\theoremstyle{thmstyletwo}%

\theoremstyle{thmstylethree}%

\begin{document}

\title[Article Title]{DrSai: an agentic AI system toward autonomous physics analysis workflows at BESIII}


\author[1,2]{\fnm{Mingfeng} \sur{He}} 
\author[1]{\fnm{Fayu} \sur{Jiang}} 
\author[3]{\fnm{Junkun} \sur{Jiao}} 
\author[1,2]{\fnm{Mingrun} \sur{Li}} 
\author*[1,2]{\fnm{Ke} \sur{Li}} \email{like@ihep.ac.cn (K.~Li)}

\author[1,2]{\fnm{Yipu} \sur{Liao}} 
\equalcont{These authors contributed equally to this work.}

\author[1,2]{\fnm{Beijiang} \sur{Liu}} 
\author[1,2]{\fnm{Tong} \sur{Liu}} 
\author[1,2]{\fnm{Fazhi} \sur{Qi}} 
\author[4,5,6]{\fnm{Zijie} \sur{Shang}} 
\author[3]{\fnm{Weimin} \sur{Song}} 
\author[1,2]{\fnm{Yue} \sur{Sun}} 
\author[1,2]{\fnm{Hong} \sur{Wang}} 
\author[4,5,6]{\fnm{Xiongfei} \sur{Wang}} 
\author[1,2]{\fnm{Dongbo} \sur{Xiong}} 
\author[1,2]{\fnm{Changzheng} \sur{Yuan}} 

\author[1,2]{\fnm{Bolun} \sur{Zhang}} 
\equalcont{These authors contributed equally to this work.}

\author*[1,2]{\fnm{Zhengde} \sur{Zhang}}\email{zdzhang@ihep.ac.cn (Z.~D.~Zhang)}
\author[7,8,9]{\fnm{Xuliang} \sur{Zhu}} 

\affil[1]{\orgname{Institute of High Energy Physics, CAS}, \orgaddress{\city{Beijing} \postcode{100049}, \country{China}}}
\affil[2]{\orgname{University of Chinese Academy of Sciences}, \orgaddress{\city{Beijing} \postcode{100049}, \country{China}}}
\affil[3]{\orgname{Jilin University}, \orgaddress{\city{Changchun} \postcode{130012}, \country{China}}}
\affil[4]{\orgname{School of Physical Science and Technology, Lanzhou University}, \orgaddress{\city{Lanzhou} \postcode{730000}, \country{China}}}
\affil[5]{\orgname{Lanzhou Center for Theoretical Physics, Key Laboratory of Theoretical Physics of Gansu Province, Key Laboratory of Quantum Theory and Applications of MoE, Gansu Provincial Research Center for Basic Disciplines of Quantum Physics, Lanzhou University}, \orgaddress{\city{Lanzhou} \postcode{730000}, \country{China}}}
\affil[6]{\orgname{Frontiers Science Center for Rare Isotopes, Lanzhou University}, \orgaddress{\city{Lanzhou} \postcode{730000}, \country{China}}}
\affil[7]{\orgname{Tsung-Dao Lee Institute, Shanghai Jiao Tong University}, \orgaddress{\city{Shanghai} \postcode{201210}, \country{China}}}
\affil[8]{\orgname{Institute of Nuclear and Particle Physics, School of Physics and Astronomy, Shanghai Jiao Tong University}, \orgaddress{\city{Shanghai} \postcode{200240}, \country{China}}}
\affil[9]{\orgname{Key Laboratory for Particle Astrophysics and Cosmology (MOE), Shanghai Key Laboratory for Particle Physics and Cosmology (SKLPPC)}, \orgaddress{\city{Shanghai} \postcode{200240}, \country{China}}}







\maketitle

High Energy Physics (HEP) experiments, such as the BEijing Spectrometer III (BESIII)~\cite{BESIII:2009fln} at the Beijing Electron-Positron Collider (BEPCII)~\cite{BESIII:2020nme}, which is a general-purpose particle detector designed to study physics in the tau-charm energy region, generate large amounts of high-dimensional and heterogeneous data. Extracting physics results from these petabyte- and exabyte-scale datasets involves complex workflows, including Monte Carlo simulation, detector objects reconstruction, signal extraction, statistical analysis, and systematic uncertainty evaluation. This knowledge-intensive process traditionally relies heavily on human experts who spend months or even years manually designing and executing analysis chains to achieve specific physics goals.

Although traditional methods have led to many significant discoveries, the emergence of Large Language Models (LLMs) offers a new way to address these challenges. With their capabilities in natural language understanding, logical reasoning, and code generation, LLMs can interpret scientific tasks and translate them into executable analysis steps. Building on these capabilities, agentic AI integrates tools and processes that allow for large-scale, systematic scans, far exceeding what can be achieved manually. By automating repetitive engineering tasks, agentic AI allows researchers to focus on more in-depth exploration. Utilizing the power of agentic AI can greatly optimize the research process, improve efficiency, and accelerate the pace of discovery while complementing human expertise. By integrating LLMs with specialized HEP tools, such as CERN ROOT~\cite{ROOT_NIMA_1997} and the BESIII Offline Software System (BOSS)~\cite{Zou:2024pmc}, we can build an ``AI partner" that conducts the full analysis either under expert supervision or automatically.

In this article, we present an implementation of this vision through DrSai~\cite{Zhang2024DrSai,Li:2026krn}, an LLM-powered multi-agent system designed for autonomous HEP analysis. By using a formalized knowledge representation scheme and ensuring alignment with scientific tools and codes, DrSai translates natural language instructions into rigorous physics analysis workflows. We validate this approach by performing large-scale re-measurements of ten different branching fractions of $J/\psi$ decays. Without manual coding, DrSai successfully navigated the real-world BESIII computing environment and produced results in excellent agreement with established benchmarks.

\begin{figure}[htbp]
    \centering
    \includegraphics[width=0.9\linewidth]{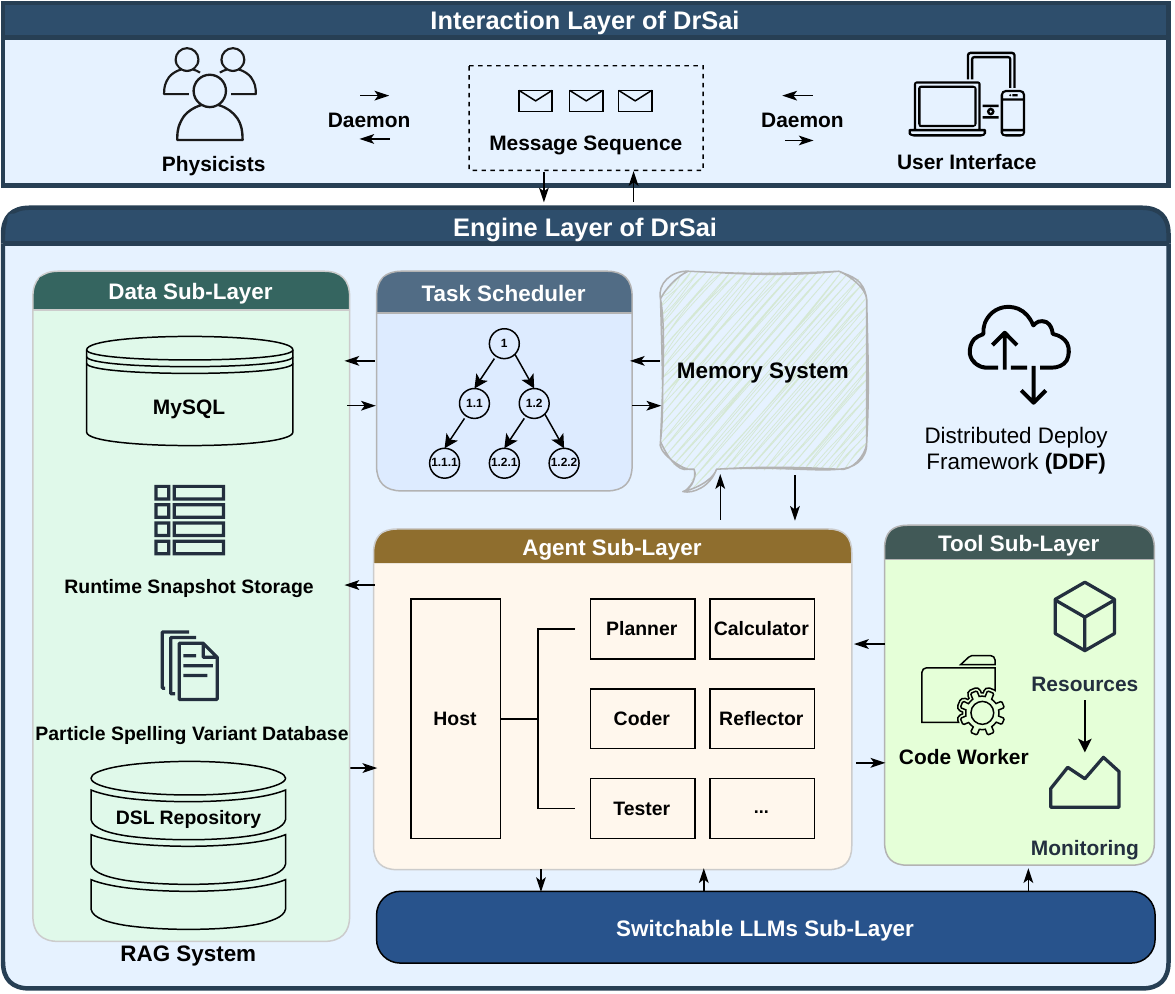}
    \caption{Architecture of the DrSai system and its multi-agent orchestration. The Host agent acts as the central controller, processing user queries to strategically dispatch tasks to specific agents within the pool. Each specialized agent is configured with LLMs and domain-specific RAG collections. A distributed deployment system manages the underlying scientific tools and high-performance computing environment. To ensure the reliability of long-running tasks and enable asynchronous re-engagement, a dedicated daemon process monitors and archives all message flows. }
    \label{fig:drsai-architecture}
\end{figure}

The DrSai system is designed to manage the complete chain of data processing in HEP research, integrating multi-agent collaboration with robust execution modules to connect human intent with physics discovery. Its architecture, illustrated in Fig.~\ref{fig:drsai-architecture}, features a central Host agent that processes user queries and dispatches tasks to specialized agents. Each agent is equipped with LLMs and domain-specific Retrieval-Augmented Generation (RAG) collections. Meanwhile, a daemon process ensures reliable communication and manages long-running tasks. This setup allows physicists to submit requests in natural language through a dialogue-based interface, triggering an automated workflow from planning to execution and visualization without manual coding.

DrSai operates within a Multi-Agent System (MAS) architecture, which supports specialized agents that work together to decompose and execute complex tasks. The system is built on the Open-DrSai framework~\cite{opendrsai}, which is based on the AutoGen framework~\cite{wu2023autogenenablingnextgenllm}, and follows a Talker-Reasoner architecture~\cite{christakopoulou2024agentsthinkingfastslow}. It employs a hierarchical management framework that organizes tasks into sub-tasks, regardless of their complexity. Each task record contains essential metadata, allowing for flexible management of research workflows. The Qwen3-max and DeepSeek-v3.2 serve as the base LLMs and provide foundational reasoning capabilities\footnote{The model choice and nominal solutions were determined on Dec. 2025.}, with RAG modules augmenting agent memory through access to external text corpora, thereby enhancing the accuracy of generated outputs. Furthermore, the integration of a server-side Remote Worker facilitates coding and execution across various programming environments in conjunction with the BOSS~\cite{Zou:2024pmc}.

To improve the generation of HEP-specific code, DrSai employs domain constraints and a configuration-driven workflow to standardize analysis components into templates. The development of \texttt{HepScript}~\cite{Jiao:2026dsu}, which builds a knowledge base of domain-specific language (DSL) from published BESIII papers, formalizes human analytical logic by translating high-level intentions into structured instructions. Additionally, a Particle Spelling Variant Database resolves inconsistencies in nomenclature, ensuring accurate retrieval during analysis. Other operational modules include runtime snapshot mechanisms for monitoring, an Anchor Word mechanism to reduce semantic ambiguity, adaptive message offloading to prevent context overflow, and a user-friendly web interface~\footnote{Deployed at \url{https://opendrsai.ihep.ac.cn}. This platform supports general-purpose agent deployment. Given the confidentiality and privacy constraints of BESIII collaboration data, access to Dr.Sai is limited exclusively to BESIII internal members.} for real-time task monitoring. Overall, the workflow begins with the LLM proposing solutions, which human experts then refine by setting key parameters. The model selects the appropriate expert agents for each task. Expert-designed code is wrapped in the DSL, accompanied by tools and instructions. The model accesses BESIII-specific data through the RAG system, automatically creates DSL scripts, and uses the tools to complete tasks efficiently. More details are available in the supplementary materials.

As a pioneering application, the DrSai system was deployed for the first time in an autonomous HEP research scenario to re-measure the branching fractions of the $J/\psi$ particle.\footnote{$J/\psi$ is a bound state made of a charm quark and an anti-charm quark. It can decay into many different light hadronic final states, such as pions, kaons, protons, and others. The branching fraction is defined as the ratio of the number of decays into a specific final state to the total number of decays of the parent particle.} Charmonium states have been studied for over fifty years, yet their decay dynamics remain ambiguous. Specifically, the ``12\% rule" test in $\psi(2S)$ and $J/\psi$ decays, defined as $Q_h = {\cal B}_{\psi(2S)\to h}/{\cal B}_{J/\psi\to h} \approx 12\%$, is influenced by contamination from continuum processes and interference between charmonium decays and continuum production~\cite{Guo:2022gkg}. The decay $\psi(2S) \to \pi^+\pi^- J/\psi$ provides a continuum-free sample and allows for a reliable check of existing results through the ``tag-and-probe" method, which mitigates interference from continuum processes that often complicate traditional $e^+e^-$ energy scan techniques. This advancement could enable BESIII to re-measure all $J/\psi$ branching fractions using $\psi(2S)$ data in the future, providing significant experimental input for perturbative QCD calculations~\cite{Brambilla:2010cs} and enhancing our understanding of charmonium decay mechanisms.

The DrSai system automates the entire BESIII analysis process, from the initial query to signal extraction and preliminary systematic uncertainty estimation. The branching fraction measurement task is decomposed into twelve sequential sub-tasks, including data pre-processing, event selection, visualization, and the final calculation of physical observables for specific decay channels. This approach highlights the capabilities of the MAS in specialized code generation and long-term process management. The current implementation primarily focuses on key contributors to systematic uncertainty, particularly reconstruction efficiencies and cited input values. More complex effects such as peaking background contributions are beyond the current scope due to the logic limitation, although a more comprehensive framework is planned for future development.

\begin{figure}[htbp]
    \centering
    \includegraphics[width=0.9\linewidth]{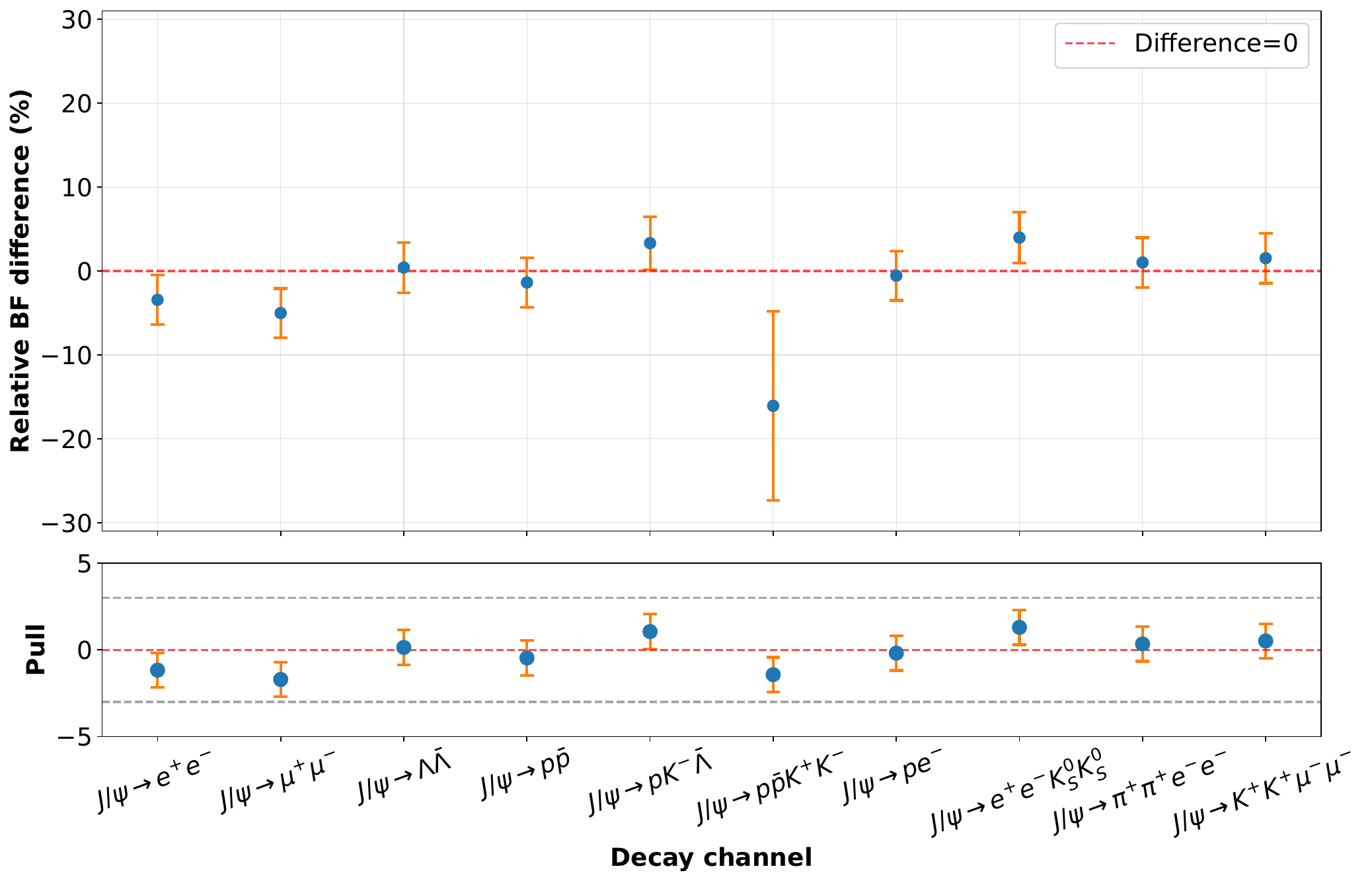}
    \caption{Validation of DrSai on ten benchmark decay channels. The relative differences between the measured and input branching fractions (BFs, denoted as ${\cal B}$) and the corresponding pull values $({\cal B}_{\rm out}-{\cal B}_{\rm in}) / \sigma_{{\cal B}_{\rm out}}$ show overall good agreement, supporting the reliability of the autonomous workflow. Detailed channel-by-channel results are given in the supplementary materials.}
    \label{fig:BR_diff_short}
\end{figure}

To assess the system, we analyzed an inclusive Monte Carlo sample with statistics equivalent to the 2009 $\psi(2S)$ dataset from BESIII, comprising over $107.7 \times 10^6$ resonant $e^+e^- \to \psi(2S)$ events and incorporating all known physical processes. The simulation utilized a \texttt{GEANT4}-based~\cite{GEANT4:2002zbu} framework and included various decay channels, testing the robustness of the system against rare decay signals and potential Beyond the Standard Model processes. A key highlight of the workflow is the fitting on the invariant mass distribution for $J/\psi$ candidates, yielding branching fractions~\footnote{The branching fraction is extracted by fitting the $J/\psi$ invariant mass distribution. From this, the branching fraction and its associated statistical uncertainty were determined using the following formula: ${\cal B}~(\sigma_{\cal B}) = N_{\rm sig}~(\sigma_{N_{\rm sig}}) /  [ N_{\psi(2S)} \times \varepsilon \times {\cal B}_{\rm sub} ]$, where $\sigma$ denotes the statistical uncertainty, $N_{\rm sig}$ is the fitted signal yield, $N_{\psi(2S)}$ is the total number of produced $\psi(2S)$ events, $\varepsilon$ is the detection efficiency estimated from signal MC samples, and ${\cal B}_{\rm sub}$ is the branching fraction for the subsequent decays.} with excellent agreement to input values~\cite{ParticleDataGroup:2024cfk}, as shown in Fig.~\ref{fig:BR_diff_short}. Additional simultaneous measurements across multiple decay channels confirmed the consistency and reliability of our system for large-scale automated physics analyses. More details can be found in the supplementary materials.

A systematic performance evaluation of the DrSai system was conducted to characterize its success rates, failure patterns, and resource consumption across several LLMs with the same configurations, including Qwen3-max, Qwen3.6-plus, GLM-4.7, GLM-5.2, DeepSeek-v3.2, DeepSeek-v4-flash, DeepSeek-R1, and GPT-4o. The evaluation dataset consisted of ten validated HEP measurement queries, with each query being sampled ten times per model, yielding 100 test samples for each LLM. The benchmark tasks followed a ``plan-first, execute-later" workflow. This rigorous multi-model assessment provides a comparative analysis of how foundational architectures handle the complex, multi-step demands of HEP research. Two primary metrics were employed for quantitative performance assessment. One metric is the overall success rate, which measures the proportion of samples that successfully complete the entire sequence of sub-tasks. The other metric is the sub-task success rate, which reflects the ratio of successfully completed sub-tasks to the total executed. For failure mode analysis, a standardized protocol was established to capture only the first occurrence of failure for each sub-task, allowing for clearer identification of root causes. Eight primary failure modes were identified, and the analysis demonstrated that failures were mainly concentrated within the tool-invocation pipeline. Certain tool-calling failure modes may bias the values used in the branching fraction calculations, causing the measured results shown in Fig.~\ref{fig:BR_diff_short} to deviate from the input values.

The system performance is quantitatively evaluated using several visualizations: success rate heatmaps, end-to-end success rate, failure mode distribution, token usage and cost profiles, and average retry depth. Ablation studies are conducted on the RAG system, DSL compilation, and particle-spelling database. Additionally, alternative prompts are tested to compare results. Figures and relevant details can be found in the supplementary materials. The success rate heatmaps (under varying retry limits) indicated that most models showed improvement as the retry budget increased, such as Qwen3-max, DeepSeek-v3.2, and GLM-5.2 achieving near-perfect success rates in most tasks. Conversely, DeepSeek-R1 exhibited only marginal improvements, primarily due to limitations in its function-calling architecture. GPT-4o, while adept at initiating tool calls, struggled with generating valid structured parameters, which resulted in persistent stallouts during execution. More advanced models, such as DeepSeek-v4-flash, may fail in structured output tasks because their reasoning processes are too complex. The token consumption analysis highlighted that models like DeepSeek-R1 faced excessive resource overhead due to ineffective tool calling, while high-performing models maintained efficient resource usage. Overall, the models with tool-calling capabilities achieve higher success rates, while more advanced models may fail on the task due to too many reasoning steps. This suggests that for some simple tasks, using more advanced models is not always necessary.

The evaluation by DrSai reveals that the main challenges in autonomous HEP research lie in execution precision and domain integration. Future efforts should focus on two key areas. First, there is a critical need for expertise in tool-calling and code generation, as current LLMs struggle to meet the requirements of HEP software systems\footnote{This statement is based on the models and system configurations tested in this study.}, often failing to produce schema-compliant parameters or error-free code for complex applications like BOSS. Thus, future models should incorporate specialized layers that understand HEP tool syntax deeply. Second, a major hurdle is the lack of formal representations for expert heuristics, which encompass the implicit knowledge experts use to handle data structures, troubleshoot fits, and ensure physical consistency. Research should aim to effectively represent and embed this domain-specific knowledge into agent architectures, transforming static information into executable logic that mirrors human decision-making.

To summarize, we have developed and validated DrSai, an autonomous multi-agent system tailored for high-energy physics analysis. By integrating LLMs with specialized HEP tools and a domain-specific knowledge base, the system successfully transitioned from natural language research objectives to physics measurements. The experimental validation at BESIII confirms that DrSai can maintain physical consistency and handle the rigorous demands of real-world experimental frameworks.

Our systematic evaluation across multiple frontier models highlights a critical insight for the AI4S (AI for Science) community: in workflow tasks with relatively fixed procedural steps and tool repositories, the core challenge for autonomous discovery is no longer the capacity for logical reasoning, but the precision of professional tool-calling or code generation and the formal representation of expert experience. While current LLMs demonstrate strong general capabilities, they frequently struggle with the ``last mile" of generating schema-compliant, executable code within complex software systems like BOSS.

The transition from a chat-based assistant to a task-driven autonomous scientist requires a deeper focus on how expert heuristics are represented and embedded within the agentic architecture. Future iterations of DrSai will prioritize the development of more sophisticated domain-synthesis layers and the expansion of the underlying physics ontology through \texttt{HepScript}-2.0~\cite{Jiao:2026dsu}. Ultimately, this research paves the way for a new era of collaborative scientific discovery, where AI agents act as competent partners in navigating the complexities of the subatomic world.

The main tasks of physical analysis at BESIII, as well as other high-energy collider experiments, are to measure branching ratios and cross sections of various processes and to study possible intermediate states. Their analysis workflows are similar to those described in this paper. Moreover, the repetitive analysis of the multi-energy data collected by BESIII inspires a strong need for automated analysis tools like DrSai. Currently, the application of the DrSai system to real physics analyses is in progress, indicating a potential shift in the approach to HEP analysis.

\backmatter


\bmhead{Conflict of interest}
The authors declare that they have no conflict of interest.

\bmhead{Acknowledgements}
This work was supported by the Strategic Priority Research Program of Chinese Academy of Sciences (XDA0480600); the 2023 CAS Outstanding Talent Program for Library and Information Science (E429C2M1); the Program of Science and Technology Development Plan of Jilin Province of China (20230101021JC); the National Natural Science Foundation of China (12247101); the Fundamental Research Funds for the Central Universities (lzujbky-2025-ytA05, lzujbky-2025-it06, lzujbky-2024-jdzx06); the Natural Science Foundation of Gansu Province (22JR5RA389, 25JRRA799), and the ``111 Center" (B20063). The authors thank BESIII Collaboration for their support on simulation datasets and software. The authors also thank the IHEP computing center for their strong support on the computing resources.

\bmhead{Author contributions}  
Beijiang Liu, Fazhi Qi, and Changzheng Yuan supervised and guided the research. Mingfeng He implemented the daemon process. Fayu Jiang carried out the literature review and data collection. Junkun Jiao, Weimin Song, Tong Liu, and Ke Li developed the domain-specific language. Yue Sun contributed to the generation of domain codes. Mingrun Li, Xuliang Zhu, and Yipu Liao standardized and integrated the HEP tools. Zijie Shang and Xiongfei Wang prepared the RAG components. Hong Wang provided support for the LLMs. Dongbo Xiong and Zhengde Zhang managed the multi-agent system and computing resources. Bolun Zhang designed and developed the agents and workflows. Ke Li and Zhengde Zhang coordinated the overall efforts. Yipu Liao and Bolun Zhang designed the experiments, carried out the tests, analyzed performance, and prepared the manuscript. All authors reviewed the final manuscript. All authors are listed in alphabetical order by last name. Yipu Liao and Bolun Zhang contributed equally to this work and share first authorship.




\clearpage

\thispagestyle{plain}

\begin{center}

    \vspace*{1.5cm}

    {\LARGE
    \setlength{\baselineskip}{1.3\baselineskip}
    \textbf{Supplementary Material} for
    ``DrSai: an agentic AI system toward autonomous physics
    analysis workflows at BESIII''
    \par
    }

    \vspace{1.5cm}

\end{center}

\vspace{1cm}

\section{The DrSai multi-agent system}

\subsection{Multi-Agent collaboration mechanism}

From a technical perspective, DrSai utilizes a Multi-Agent System (MAS) architecture~\cite{han2025llmmultiagentsystemschallenges, xi2023risepotentiallargelanguage} integrated with a daemon-based message management system. The MAS consists of specialized agents that collaborate to decompose and execute complex tasks, supported by base LLMs for foundational reasoning and natural language processing. To incorporate domain-specific expertise, Retrieval-Augmented Generation (RAG)~\cite{lewis2021retrievalaugmentedgenerationknowledgeintensivenlp} modules are utilized to feed external knowledge directly into the agents' decision-making and code-generation processes. The operations are orchestrated by a central controller that assigns tasks to specialized remote workers. A daemon process manages message sequences between all components, ensuring an orderly information flow and coordinated task progress. This setup allows physicists to submit requests in natural language through a dialogue-based interface, triggering an automated workflow--from initial planning to execution and visualization--enabling a seamless experience without the need for manual coding.

An agent in this framework is an autonomous computational entity that leverages an LLM to perceive inputs, reason toward predefined goals, and perform actions using specialized tools. DrSai is built on the Open-DrSai framework~\cite{opendrsai}, which is based on the AutoGen framework~\cite{wu2023autogenenablingnextgenllm}, and follows a Talker-Reasoner architecture~\cite{christakopoulou2024agentsthinkingfastslow}. The Host agent serves as the primary entry point, inferring user intent to decide whether to respond independently or coordinate expert agents. This architecture constrains agent behaviors, improves response accuracy, and supports dynamic prompt adjustments to maintain a consistent system identity. Every interaction produces a ``Task-Conclusion" message, ensuring transparent session management and clear tracking of progress.

In this study, the DrSai system comprises six agents:
\begin{itemize}
    \item \textbf{Host}: Routes user requests and manages cross-agent collaboration.
    \item \textbf{Planner}: Decomposes and structures complex domain tasks.
    \item \textbf{Coder}: Generates codes and maps key parameters using internal tools.
    \item \textbf{Tester}: Executes codes with remote workers and retrieves the results.
    \item \textbf{Calculator}: Computes branching fractions of the decay chains.
    \item \textbf{Reflector}: Validates result rationality against task context.
\end{itemize}

To mitigate the risk of disconnection during long-term processing on remote clusters (e.g., via HTCondor~\cite{htcondor}), we implemented a communication daemon. This daemon maintains a persistent connection with the MAS while providing the front-end with decoupled, reusable access. It handles message routing, job termination, and caching, allowing the front-end to reconnect seamlessly and manage resources efficiently.

\subsection{Hierarchical task system}

DrSai employs a hierarchical task management system inspired by fractal-like, self-similar structures. Under this framework, every task--regardless of duration or complexity--follows the same organizational rules. Large-scale research objectives are decomposed into a one-to-many hierarchy of sub-tasks, simplifying the management of complex workflows and ensuring dependency tracking. Each task record contains essential metadata, including a unique ID, timestamp, initiator, and real-time status (e.g., queued, in progress, completed). This architecture ensures a flexible and reliable management layer capable of handling the intricacies of real-world HEP research.

\subsection{Language models and RAG Integration}

In this study, the Qwen3-max-2025-09-23 model~\cite{yang2025qwen3technicalreport} and the DeepSeek-v3.2 model~\cite{deepseekai2024deepseekv3technicalreport} are used as base LLMs for agents. We select specific models based on the nature of the task:
\begin{itemize}
\item Qwen3-max is primarily used for agents requiring complex reasoning and active tool invocation, as it demonstrates a stronger proactivity in function calling.
\item DeepSeek-v3.2 is preferred for straightforward, prompt-based tasks due to its superior inference speed and high accuracy.
\end{itemize}
Most agents are equipped with custom tools and integrated with RAG modules. More detailed agent configurations can be found in the Sec.~\ref{app:agent_config}.

A specialized RAG module, HEP-RAG, is integrated into the system using Llama-index~\cite{Liu_LlamaIndex_2022} and the Qdrant~\cite{Qdrant} vector database. HEP-RAG augments the agents' parametric memory by providing access to an external text corpus. For the Coder agent, specifically, multiple RAG collections are provided to enhance its ability to generate precise and executable BESIII-specific analysis scripts, significantly reducing hallucinations in domain-specific tasks.

\subsection{Remote worker and BOSS integration}

To bridge the gap between AI reasoning and physical execution, a server-side Remote Worker is implemented based on HaiDDF. Running on the user--s work server or a local cluster, the worker supports standard programming environments including \texttt{C++}, \texttt{Python}, \texttt{Shell}, and CERN ROOT~\cite{ROOT_NIMA_1997}.

Central to this environment is the BESIII Offline Software System (BOSS)~\cite{Zou:2024pmc}, the integrated system for simulation, reconstruction, and analysis at BESIII. Within DrSai, a specialized BOSS code worker communicates directly with the Tester agent. Upon receiving execution commands--such as compiling an analysis algorithm--the worker performs the task on the server and returns the results, including logs and error information. This completes the transformation of DrSai from a conversational assistant into an instruction-driven execution system.

\subsection{Domain constraints and knowledge formalization}

While LLMs excel in general-purpose tasks, their ability to generate HEP-specific code is often limited by the high complexity of domain software. To bridge this gap, we implemented a series of domain constraints designed to narrow the search space for the agents. These constraints significantly refine agent outputs, thereby enhancing the overall success rate and stability of the system. 

\paragraph{Configuration-driven code generation}
Generating BESIII analysis code from scratch is often prohibitively complex for LLMs. DrSai utilizes a configuration-driven workflow where analysis components are standardized into reusable templates. The agent provides a high-level \texttt{JSON} configuration specifying parameters like decay channels and selection criteria. The system then automatically populates these into templates to produce ready-to-run \texttt{C++} algorithms. This reduces the technical burden and ensures the generated code adheres to experimental standards.

In detail, we created these templates ourselves and assigned certain key intermediate variables to the LLM, such as the number of particles involved and the selection criteria for each particle. However, this does not mean our code is controlled only by flags; instead, the LLM is required to make decisions at multiple steps throughout the workflow. Since some BESIII code is closed-source and not publicly available due to collaboration policies, we adopted the mapping method described in the main text. In this approach, we represent many parameters as \texttt{JSON} text, which is then processed by our custom mapping functions. Within this framework, the code generated by the LLM essentially takes the form of \texttt{JSON}, greatly reducing the complexity of code generation.

\paragraph{Formalized analysis logic: \texttt{HepScript}}
A significant challenge in HEP automation is that expert analytical strategies are often rooted in heuristic experience rather than explicit documentation. We developed \texttt{HepScript}~\cite{Jiao:2026dsu}, a domain-specific language (DSL) designed to formalize human analytical logic. \texttt{HepScript} acts as a structured knowledge representation that defines the ``grammar" of physics analysis at BESIII. It abstracts the analysis process into rule-based instructions, translating high-level physical intent into standardized executable sequences. This ensures the rigor of the analysis and creates a high-quality data foundation for training future agents.

Specifically, we have built a knowledge base of DSLs extracted from all published BESIII papers. When a user performs a new analysis using DrSai, the system first retrieves the most similar DSLs from this knowledge base and then generates a candidate DSL tailored to the new task. After the user reviews and optionally modifies the DSL, or once the full workflow is completed and verified, the updated DSL is incorporated back into the knowledge base, thereby capturing expert feedback in a formal, reusable form. In this work, we have established a very preliminary version of this module. The main challenge we are currently addressing is how to accurately represent and integrate the expert modifications and feedback into the DSL generation process, so that the system can effectively learn from human corrections. We believe this iterative refinement will eventually provide a practical solution for encoding heuristics, and we are actively working on this part~\cite{Jiao:2026dsu}.

\paragraph{Particle spelling variant database}
To resolve inconsistencies in particle nomenclature, we developed a Particle Spelling Variant Database. While tools like \texttt{PyHEP}~\cite{beringer2024ichep}, \texttt{PDG}~\cite{pdg2023hadron}, and \texttt{Particle}~\cite{particle_pypi} provide authoritative catalogs, they often lack informal spelling variants found in user queries. Our database unifies diverse spellings under a unique MC Identifier, ensuring accurate mapping and attribute retrieval for every particle during decay chain parsing.

\subsection{System operational modules}

\paragraph{Runtime snapshot and data logging}
DrSai incorporates an automated mechanism that captures runtime snapshots of agent interactions, reasoning flows, and tool usage. These execution traces enable real-time monitoring and rapid diagnosis, while providing a dataset for historical analysis and future model fine-tuning.

\paragraph{Anchor word mechanism}
To resolve semantic ambiguity in specialized HEP tasks, we introduced an Anchor Word mechanism. By embedding expert-defined keywords into agent descriptions, we guide the LLM to precisely match user requirements with the appropriate expert agents, stabilizing task decomposition.

\paragraph{Adaptive message offloading}
To prevent context overflow, we implemented a message offloading strategy. The Host agent maintains the complete dialogue history for coordination, while specialized agents receive only the minimal context necessary for their current task stage. Critical parameters (e.g., file paths, physical constraints) are preserved as global state information to reduce redundancy.

\paragraph{User interface and monitoring}
A web-based interface using the OpenWebUI framework~\cite{OpenWebUI} provides near real-time monitoring of task progress~\footnote{Deployed at \url{https://opendrsai.ihep.ac.cn}. This platform supports general-purpose agent deployment. Given the confidentiality and privacy constraints of BESIII collaboration data, access to Dr.Sai is limited exclusively to BESIII internal members.}. By moderately compressing interaction logs, the interface generates concise status updates, allowing physicists to oversee the Research process effectively.

\section{Re-measurement of $J/\psi$ branching fractions by DrSai}\label{sec:bf}  

As a pioneering application, the DrSai system was deployed for the first time in a real-world HEP research scenario. Charmonium states have been observed and studied for over fifty years; however, their decay dynamics remain unclear. The ``12\% rule" test in $\psi(2S)$ and $J/\psi$ decays, defined as $Q_h = {\cal B}_{\psi(2S)\to h}/{\cal B}_{J/\psi\to h} \approx 12\%$, is affected by contamination from continuum processes and interference between charmonium decays and continuum production~\cite{Guo:2022gkg,BESIII:2023syz}. The $\psi(2S) \to \pi^+\pi^- J/\psi$ decay provides an independent, continuum-free sample. Using the ``tag-and-probe" method effectively reduces interference from continuum processes that is common in traditional $e^+e^-$ energy scan techniques, offering a reliable and independent check of existing results. BESIII can re-measure all the $J/\psi$ branching fractions using $\psi(2S)$ data in the future. Such measurements supply important experimental input for perturbative QCD calculations~\cite{Appelquist:1974zd,DeRujula:1974rkb,Brambilla:2010cs} and improve our understanding of charmonium decay mechanisms. In this work, we perform the first test using MC samples.

Series of videos or GIF files showing the full or failed workflows can be found in DrSai example (\href{https://github.com/waxiliwwww/drsai_related}{https://github.com/waxiliwwww/drsai\_related}).

\subsection{Task understanding and decomposition}
\begin{figure}[htbp]
    \centering
    \includegraphics[width=0.7\linewidth]{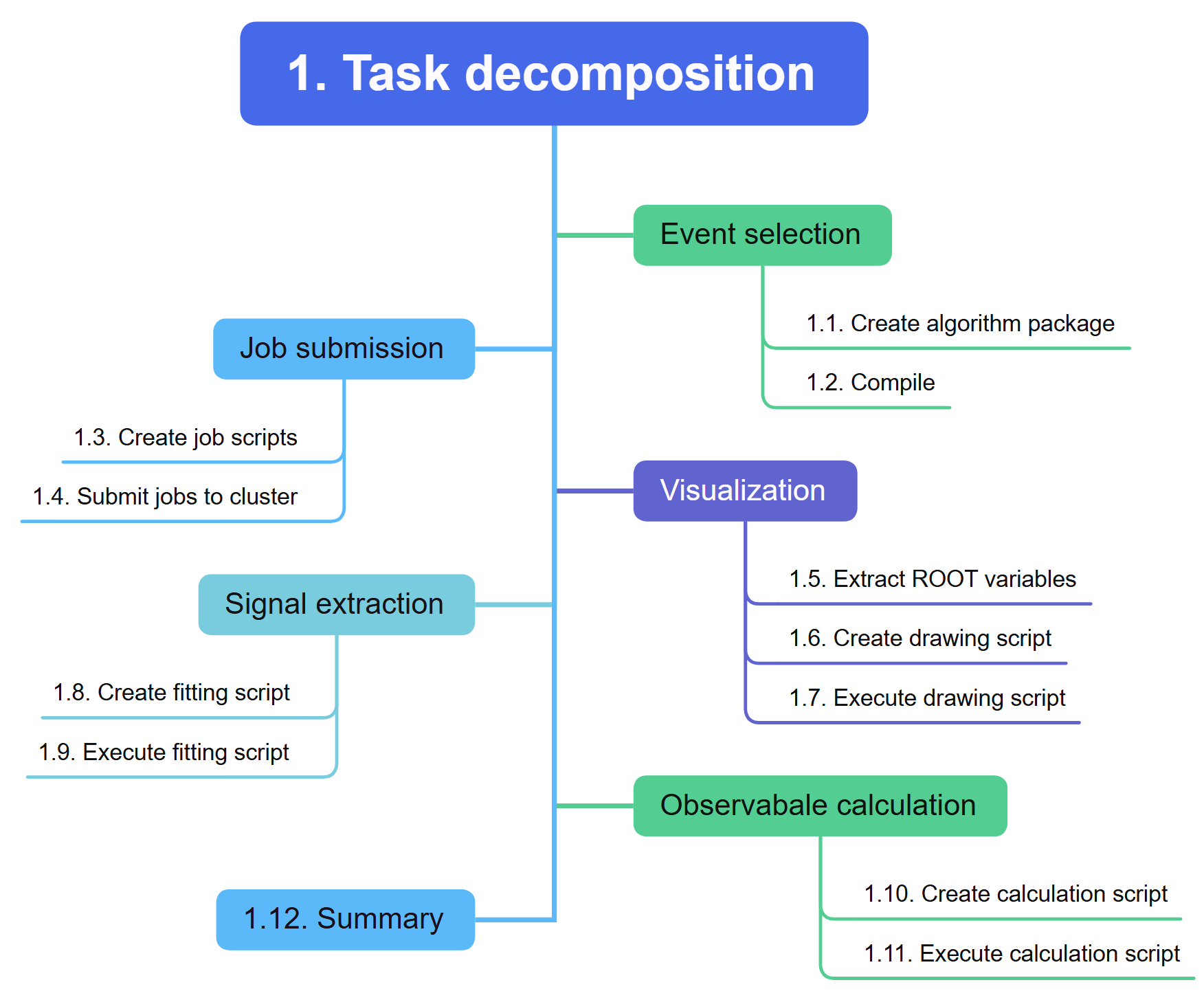}
    \caption{The physical analysis workflow at BESIII.} 
    \label{fig:BESIII_workflow}
\end{figure}

The DrSai system automates the complete BESIII analysis chain from the initial natural language query to signal extraction and preliminary systematic uncertainty estimation, as illustrated in Fig.~\ref{fig:BESIII_workflow}.

The complex task of branching fraction measurement is decomposed into twelve sequential sub-tasks. These encompass data pre-processing, event selection, job management, data parsing, visualization, signal extraction, and the final calculation of physical observables for specific decay channels. The execution of these sub-tasks demonstrates the core capabilities of the MAS, particularly in specialized code generation, long-term process management, and maintaining logical consistency across a complex analysis pipeline.

Due to the inherent complexity of systematic uncertainty estimation, the current implementation focuses on primary contributors. In this study, we consider uncertainties arising from reconstruction efficiencies and cited input values. More complex effects such as peaking background contributions are beyond the current scope due to the logic limitation, although a more comprehensive framework is planned for future development. In actual analyses, dealing with some uncertainties may even be better treated as an independent study. Table~\ref{tab:syst_uncer} lists the comparisons between the publication-level systematic uncertainties and the level achievable in this work.

\begin{table}[htbp]
  \centering
  \caption{Comparisons between the publication-level systematic uncertainties and the level achievable in this work.}
  \begin{tabular}{lcc}
    \toprule
    Source of systematic uncertainty & Publication-level & DrSai (This work) \\ \midrule
    Luminosity / $N_{\psi(2S)}$ & \yes & \yes \\
    Tracking / PID efficiency & \yes & \yes \\
    Neutral reconstruction efficiency & \yes & \yes \\
    Unknown MC modeling & \yes & \no \\
    Selection-induced bias & \yes & \no \\
    MC statistics & \yes & \yes \\
    Cited branching fraction & \yes & \yes \\
    Fit methods & \yes & \no \\
    Peaking background & \yes & \no \\
    \bottomrule
  \end{tabular}
  \label{tab:syst_uncer}
\end{table}


\subsection{Simulation samples}

To evaluate the system, we analyzed an inclusive MC sample simulated via BOSS~\cite{Zou:2024pmc}, with statistics equivalent to the 2009 $\psi(2S)$ dataset collected at BESIII. This sample comprises $107.7 \times 10^6$ resonant $e^+e^-\to\psi(2S)$ events~\cite{BESIII:2024lks}, alongside initial-state radiation production of the $J/\psi$ and other continuum processes.

The simulation utilizes a \texttt{geant4}-based~\cite{GEANT4:2002zbu} framework incorporating the full geometric description and detector response of BESIII. Beam energy spread and initial-state radiation effects are modeled using the \texttt{kkmc} generator~\cite{Jadach:1999vf,Jadach:2000ir}. Particle decays are managed by \texttt{evtgen}~\cite{Lange:2001uf,Ping:2008zz}, with branching fractions sourced from the Particle Data Group (PDG)~\cite{ParticleDataGroup:2024cfk} or estimated via \texttt{lundcharm}~\cite{Chen:2000tv,Yang:2014vra} where experimental data are unavailable. Final-state radiation from charged particles is simulated using \texttt{photos}~\cite{Barberio:1990ms}.

The reference branching fractions for the targeted final states are determined by aggregating all intermediate states within the MC samples, as detailed in Table~\ref{tab:bench_BR}. Beyond standard decay modes, we intentionally included rare decay signals and potential Beyond the Standard Model processes to test the mixed branching fraction calculations. These diverse channels provide a rigorous test of the system's robustness and sensitivity. In addition to the inclusive MC, dedicated signal MC samples were generated to determine detection efficiencies--a critical parameter for the automated branching fraction measurements discussed in the subsequent sections.

\begin{table}[htbp]
  \centering
  \caption{Summary of benchmark decay channels used in branching fraction measurement. The last two lines give the measured branching fractions for final states with $\pi^+\pi^-$. These measurements suffer from biased efficiency estimates caused by background contamination. ${\cal B}_{\rm in}$ is the input branching fraction, and ${\cal B}_{\rm meas}$ is the measured branching fraction where the first error is statistical and the second is systematic. BNV means the process with Baryon number violation, and CLFV means the process with charged lepton number violation. The asterisk on the branching fraction indicates that its value is obtained by including the mixed signal yield.}
  \begin{tabular}{lccc}
    \toprule
    Decay Channel & ${\cal B}_{\rm in}$ ($\times10^{-3}$) & ${\cal B}_{\rm meas}$ ($\times10^{-3}$) & Comment \\
    \midrule
    $\psi(2S) \rightarrow \pi^+ \pi^- [J/\psi \rightarrow e^+ e^-]$ & 59.71 & $57.73\pm0.68\pm1.55$ & - \\
    $\psi(2S) \rightarrow \pi^+ \pi^- [J/\psi \rightarrow \mu^+ \mu^-]$ & 59.30 & $56.47\pm0.66\pm1.52$ & - \\
    $\psi(2S) \rightarrow \pi^+ \pi^- [J/\psi \rightarrow \Lambda\bar{\Lambda}]$ & 1.89 & $1.898\pm0.025\pm0.051$ & - \\
    $\psi(2S) \rightarrow \pi^+ \pi^- [J/\psi \rightarrow p\bar{p}]$ & 2.12 & $2.091\pm0.025\pm0.056$ & - \\
    $\psi(2S) \rightarrow \pi^+ \pi^- [J/\psi \rightarrow p K^{-}\bar{\Lambda}]$ & 0.45 & $0.460\pm0.008\pm0.012$ & - \\
    $\psi(2S) \rightarrow \pi^+ \pi^- [J/\psi \rightarrow p\bar{p}K^{+}K^{-}]$ & 0.026 & $0.0220\pm0.0024\pm0.0006$ & - \\
    $\psi(2S) \rightarrow \pi^+ \pi^- [J/\psi \rightarrow p e^{-}]$ & 2.67* & $2.655\pm0.032\pm0.072$ & BNV \& CLFV \\
    $\psi(2S) \rightarrow \pi^+ \pi^- [J/\psi \rightarrow e^+e^-K^{0}_{S}K^{0}_{S}]$ & 2.67* & $2.780\pm0.041\pm0.075$ & Dalitz decay \\
    $\psi(2S) \rightarrow \pi^+ \pi^- [J/\psi \rightarrow \pi^{+}\pi^{+}e^{-}e^{-}]$ & 2.67* & $2.697\pm0.034\pm0.073$ & CLFV \\
    $\psi(2S) \rightarrow \pi^+ \pi^- [J/\psi \rightarrow K^+K^+\mu^-\mu^-]$ & 2.67* & $2.711\pm0.035\pm0.073$ & CLFV \\
    \midrule
    $\psi(2S) \rightarrow \pi^+ \pi^- [J/\psi \rightarrow \pi^+ \pi^- \pi^0]$ & 20.70 & $19.01\pm0.36\pm1.11$ & $\pi^+\pi^-$ final states \\
    $\psi(2S) \rightarrow \pi^+ \pi^- [J/\psi \rightarrow \pi^+ \pi^- \eta]$ & 0.638 & $0.986\pm0.046\pm0.060$ & $\pi^+\pi^-$ final states \\
    \bottomrule
  \end{tabular}
  \label{tab:bench_BR}
\end{table}

\subsection{Execution workflow and analysis}
Using the decay chain $\psi(2S) \rightarrow \pi^+\pi^- J/\psi~(\rightarrow e^+e^-)$ as a benchmark, the execution workflow within the DrSai system is summarized below:

\begin{description}
    \item[Event Selection:] The process begins with the user providing high-level information, such as the decay chain and desired physics observables, via a natural language interface. The Host agent evaluates this input and extracts into a global context. This triggers a step-by-step workflow led by the Coder agent; first it synthesizes C++ analysis algorithms within the BOSS by populating standardized templates with extra information of \texttt{HepScript} from RAG module; second, it identifies paths of different datasets and generates signal MC samples using BOSS-integrated generators, automatically ensuring that conservation laws (e.g., quantum numbers) are maintained. The complete package of algorithms, configurations, and job scripts is then transmitted to the Tester agent, which coordinates the storage and compilation on a remote HPC cluster via the controller. This configuration-driven approach effectively decouples high-level physics intent from low-level software implementation, significantly improving the system's generation speed and success rate.\hfill \\
    
    \item[Execution and monitoring:] Given the long-running feature of HEP jobs, the Tester agent submits scripts in batches to the HTCondor scheduler~\cite{htcondor} and enters a ``sleep mode" to optimize system resources. A specialized daemon process handles real-time monitoring, employing a gradient-based strategy to dynamically adjust query intervals. Once tasks (typically a few hours) are complete, the system automatically resumes the multi-agent session. Users can also oversee progress and trigger subsequent steps via a ``continue" command on the web interface, ensuring seamless workflow continuity. \hfill \\
    
    \item[Optimization and Visualization:] Upon job completion, the Tester agent parses the \texttt{ROOT} TTree structures~\cite{ROOT_NIMA_1997} and broadcasts the metadata to the Host agents. The Coder agent then identifies key variables and generates plotting scripts to visualize signal versus background distributions. Simultaneously, the Tester agent applies optimized event-level selection criteria to suppress background events and refine the data sample.\hfill \\

    \item[Signal Extraction:] The Coder agent synthesizes a \texttt{ROOT}-based fitting script, modeling the signal as the signal MC shape convolved with a Gaussian function to account for detector resolution. The background is described by a high-order Chebyshev polynomial. To ensure robust convergence, the system utilizes multiple random seed resets during the fitting process. The Tester agent executes the fit and reports the finalized parameters back to the message stream. The fit results and plots could be found in Sec.~\ref{app:app_fit}.\hfill \\

    \item[Calculation and Uncertainty Estimation:] Using formulas and physical constants retrieved from \texttt{HepScript}, the Coder agent generates \texttt{Python} scripts to calculate final branching fractions. The system automatically accounts for statistical errors from the fit and evaluates common BESIII systematic uncertainties based on the specific particle characteristics of the decay chain. By applying standard error propagation, the Tester agent executes the calculation to produce the final nominal value along with its total estimated uncertainty.\hfill \\
    
\end{description}

The signal yield was extracted by fitting the $J/\psi$ invariant mass distribution, from which the branching fraction and its associated statistical uncertainty were determined using the following formula:

\begin{equation}
    {\cal B}~(\sigma_{\cal B}) = \frac{ N_{\rm sig}~(\sigma_{N_{\rm sig}}) }{ N_{\psi(2S)} \times \varepsilon \times {\cal B}_{\rm sub} },
\end{equation}
where $\sigma$ denotes the statistical uncertainty, $N_{\rm sig}$ is the fitted signal yield, $N_{\psi(2S)}$ is the total number of produced $\psi(2S)$ events~\cite{BESIII:2024lks}, $\varepsilon$ is the detection efficiency estimated from signal MC samples, and ${\cal B}_{\rm sub}$ is the branching fraction for the subsequent decays. The analysis incorporates common BESIII systematic uncertainties, including those from charged-track reconstruction, particle identification~\cite{Yuan:2015wga,Liu:2024uno,Chai:2025xni}, and the total used $N_{\psi(2S)}$ count~\cite{BESIII:2024lks}. The resulting measurement shows excellent agreement with the input values used in the inclusive MC simulation.

To further evaluate the system--s performance, we performed simultaneous measurements across ten distinct decay channels, covering the primary $J/\psi$ decay modes at BESIII. Channels containing an additional $\pi^+\pi^-$ pair in the final state were excluded to avoid high mis-combination rates with the transition pions from $\psi(2S)$, a complexity currently beyond the system's logic. All channels were processed using the automated methodology described above. The consistency between measured and input values was verified through the relative difference $\delta = ( {\cal B}_{\rm out}-{\cal B}_{\rm in}) / {\cal B}_{\rm in}$ and the pull distributions $({\cal B}_{\rm out}-{\cal B}_{\rm in}) / \sigma_{{\cal B}_{\rm out}}$. This high-degree of agreement across multiple channels validates the robustness and reliability of the DrSai system for large-scale, automated physics analysis.

\newpage
\section{Performance study}  

A systematic performance evaluation of the DrSai system was conducted to characterize its success rates, failure patterns, and resource consumption across several LLMs. The evaluated models include Qwen3-max-2025-09 (Qwen3-max)~\cite{yang2025qwen3technicalreport}, DeepSeek-v3.2~\cite{deepseekai2024deepseekv3technicalreport}, GLM-4.7~\cite{5team2025glm45agenticreasoningcoding}, DeepSeek-R1~\cite{Guo_2025}, and GPT-4o~\cite{openai2024gpt4ocard}. Some more recent models, DeepSeek-v4-flash, GLM-5.2 and Qwen3.6-plus, are added in the evaluation during the journal review.

The evaluation dataset consists of ten validated HEP measurement queries. To ensure statistical reliability, each query was sampled ten times per model, yielding 100 test samples for each LLM. All benchmark tasks followed a ``plan-first, execute-later" workflow, with the tasks listed in Table~\ref{tab:bench_hep_tasks} and the workflow shown in Fig.~\ref{fig:BESIII_workflow}. This rigorous multi-model assessment provides a comparative analysis of how foundational architectures handle the complex, multi-step demands of HEP research.

\begin{table}[htbp]
  \centering
  \caption{Benchmark of high energy physics measurement tasks for the performance evaluation in the DrSai system. Each line shows the decay channel and its corresponding generation energy, along with the observable used for display.}
  \begin{tabular}{lcc}
    \toprule
    Decay Channel & Observable & Energy (GeV) \\
    \midrule
    $\psi(4260) \to \pi^+ \pi^- [J/\psi \to \mu^+ \mu^-]$ & Invariant mass of $J/\psi~(\to\mu^+\mu^-)$ & 4.260 \\
    $\psi(4260) \to K^+ K^- [J/\psi \to e^+ e^-]$ & Invariant mass of $J/\psi~(\to e^+e^-)$ & 4.260 \\
    $J/\psi \to [\rho^+ \to \pi^+ \pi^0] \pi^-$ & Invariant mass of $\rho^+~(\to\pi^+\pi^0)$ & 3.097 \\
    $J/\psi \to \eta [\phi \to K^+ K^-]$ & Momentum distribution of $\eta~(\to\gamma\gamma)$ & 3.097 \\
    $J/\psi \to \mu^+ \mu^-$ & MUC hit depth of $\mu^+$ & 3.097 \\
    $\psi(2S) \to K^0_S K^+ \pi^-$ & Momentum of $K^+$ & 3.686 \\
    $J/\psi \to p \bar{p} K^+ K^-$ & $\cos\theta$ distribution of $K^+$ & 3.097 \\
    $J/\psi \to \bar{p} K^+ \Lambda$ & $\cos\theta$ distribution of $\Lambda~(\to p\pi^-)$ & 3.097 \\
    $\psi(2S) \to \pi^+ \pi^- [J/\psi \to p \bar{p} \eta]$ & Invariant mass distribution of $J/\psi~(\to p \bar{p} \eta)$ & 3.686 \\
    $\psi(2S) \to \pi^+ \pi^- [J/\psi \to \pi^+ \pi^- \pi^0]$ & Energy distribution of $\pi^0~(\to\gamma\gamma)$ & 3.686 \\
    \bottomrule
  \end{tabular}
  \label{tab:bench_hep_tasks}
\end{table}

\subsection{Evaluation methodology}

In our framework, the end-to-end HEP measurement task is decomposed into twelve domain-specific sub-tasks, preceded by a top-level task planning step. To account for the iterative nature of problem-solving, each sub-task is permitted up to nine reflection retries to resolve errors and refine outputs.

We employ two primary metrics for quantitative performance assessment:
\begin{itemize}
    \item \textbf{Overall Success Rate}: The proportion of samples that successfully complete the entire sequence of sub-tasks. The overall success rate is defined as the proportion of cases where both the distribution plot and the fitting plot were successfully created.

    \item \textbf{Sub-task Success Rate}:  The ratio of successfully completed sub-tasks within the maximum allowed attempts to the total number of sub-tasks executed. The success rates of sub-tasks are evaluated by checking key message fields and verifying structured arguments within the agent responses.
\end{itemize}

For failure mode analysis, we implemented a standardized statistical protocol: if a sub-task fails repeatedly, only the first observed failure is recorded for that instance. By focusing on the initial error before subsequent retries, this approach enables us to pinpoint the root cause of failure more precisely. This methodology ensures a consistent and reliable comparison of workflow robustness across different models and experimental rounds, preventing secondary errors (induced by the reflection process itself) from masking the primary limitations of the models.

By cross-referencing key message fields and validating structured arguments within agent responses, we identified and classified eight primary failure modes:

\begin{itemize}
    \item \textbf{Speaker mismatch}: The agent name in the message deviates from the expected role. This error typically stems from ambiguous role definitions, the Host agent's misjudgment of the dialogue state, or inherent limitations in the model's instruction-following capabilities.
    
    \item \textbf{Tool Call Missing}: The agent fails to invoke a required tool to resolve a sub-task, providing irrelevant natural language or halting the workflow instead. This usually points to a failure in contextual grounding or the model--s inability to recognize the necessity of an external function call.

    \item \textbf{Tool Call Mismatch}: The agent invokes an incorrect tool, either by hallucinating a non-existent function name or selecting a sub-optimal tool from the available registry. This reflects cognitive confusion within complex contexts or limitations of the LLMs.

    \item \textbf{Spurious Tool Call}: The agent erroneously initiates a tool call when the task requires only reasoning or natural language synthesis. Similar to mismatches, this indicates an over-reliance on functional triggering or a misunderstanding of the task's current state.

    \item \textbf{Structural Inconsistency}: Although a tool is called, the generated arguments or output messages fail to adhere to the required schema (e.g., missing mandatory keys in \texttt{JSON} configurations for HEP fitting). This highlights a lack of precision in generating structured data under domain-specific constraints.

    \item \textbf{Invalid Variables}: A system-level error where variable names defined in the visualization or analysis configurations do not exist within the \texttt{ROOT} TTree branches. This signifies a failure in maintaining state consistency and coordinating results across multiple task stages.

    \item \textbf{Planning Formalization Failure}: A Planner-specific error where the output lacks the system-mandated structured format (e.g., a parsable table) required for task decomposition, violating preset processability rules.

    \item \textbf{Refinement Exhaustion}: The agent reaches the maximum limit of nine reflection attempts without achieving a successful output. This often indicates a ``reasoning loop" where the model fails to break out of a specific error pattern despite repeated feedback.

\end{itemize}

The system performance is quantitatively analyzed through some visualizations: success rate heatmaps, an overall success rate, a failure mode distribution, a token consumption and cost profile, and average retry depth distribution. These figures respectively characterize sub-task-level reliability, model-specific vulnerabilities, and the computational efficiency of each LLM during execution. We also perform ablation studies for the RAG system, the DSL compilation, and the particle-spelling database. An alternative result with different prompts are also tested.

\subsection{Success rates}

\begin{figure}[htbp]
    \centering
    \includegraphics[width=1.05\linewidth]{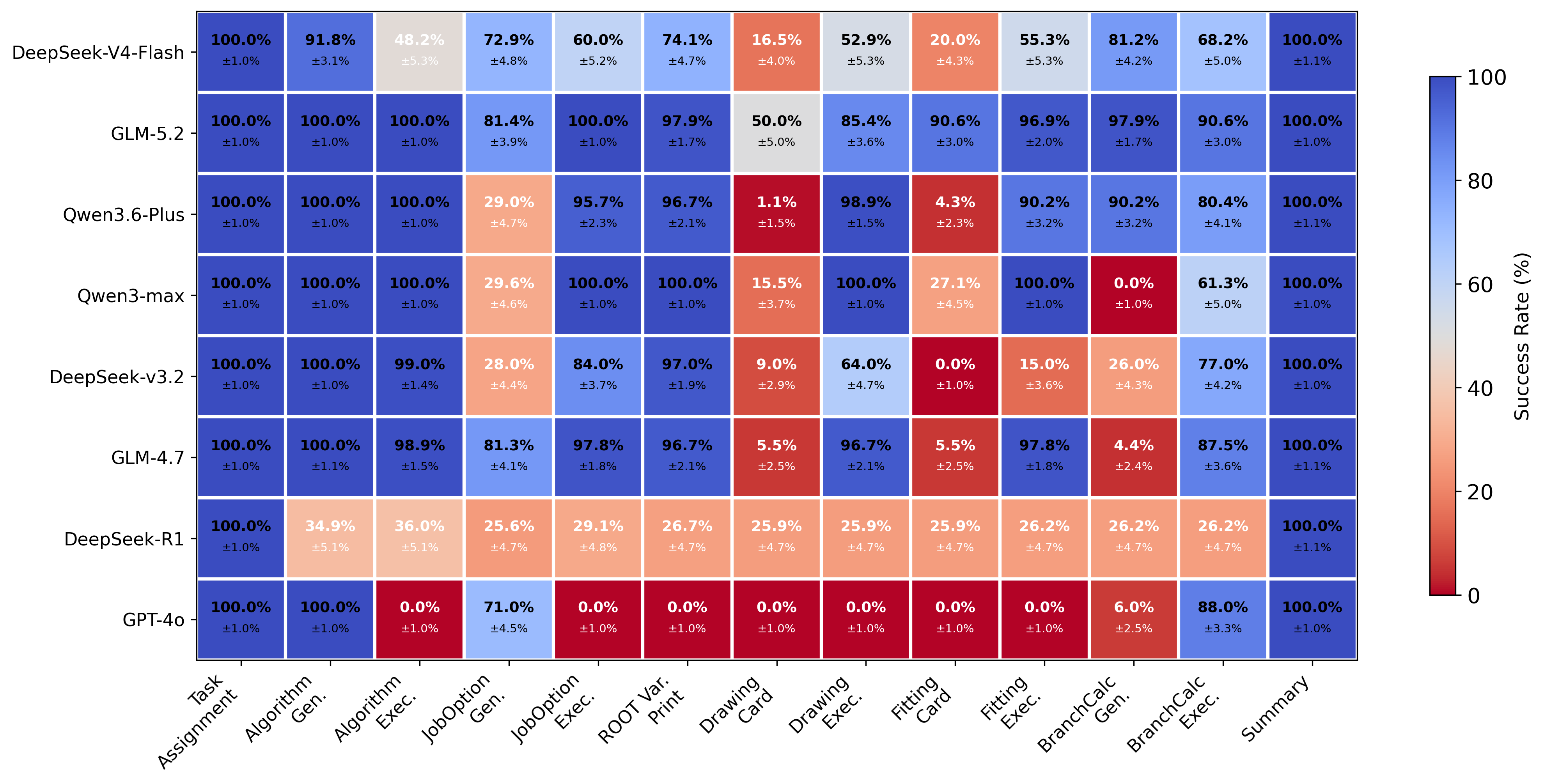}
    \caption{Success rate heatmap of different LLMs across sub-tasks under one maximum retry attempt.}
    \label{fig:max_retries_1_SR}
\end{figure}

\begin{figure}[htbp]
    \centering
    \includegraphics[width=1.05\linewidth]{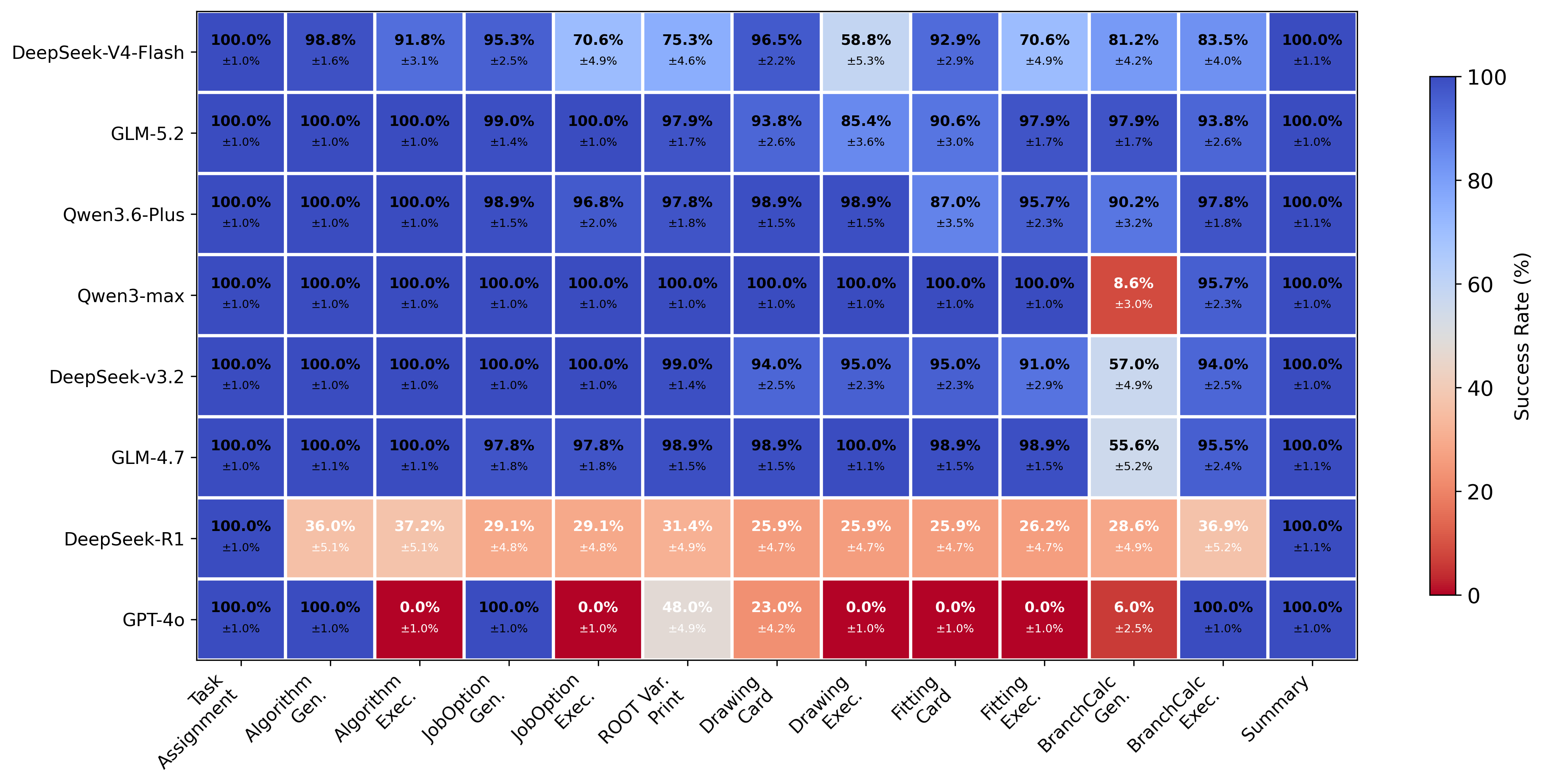}
    \caption{Success rate heatmap of different LLMs across sub-tasks under five maximum retry attempts.}
    \label{fig:max_retries_5_SR}
\end{figure}

\begin{figure}[htbp]
    \centering
    \includegraphics[width=1.05\linewidth]{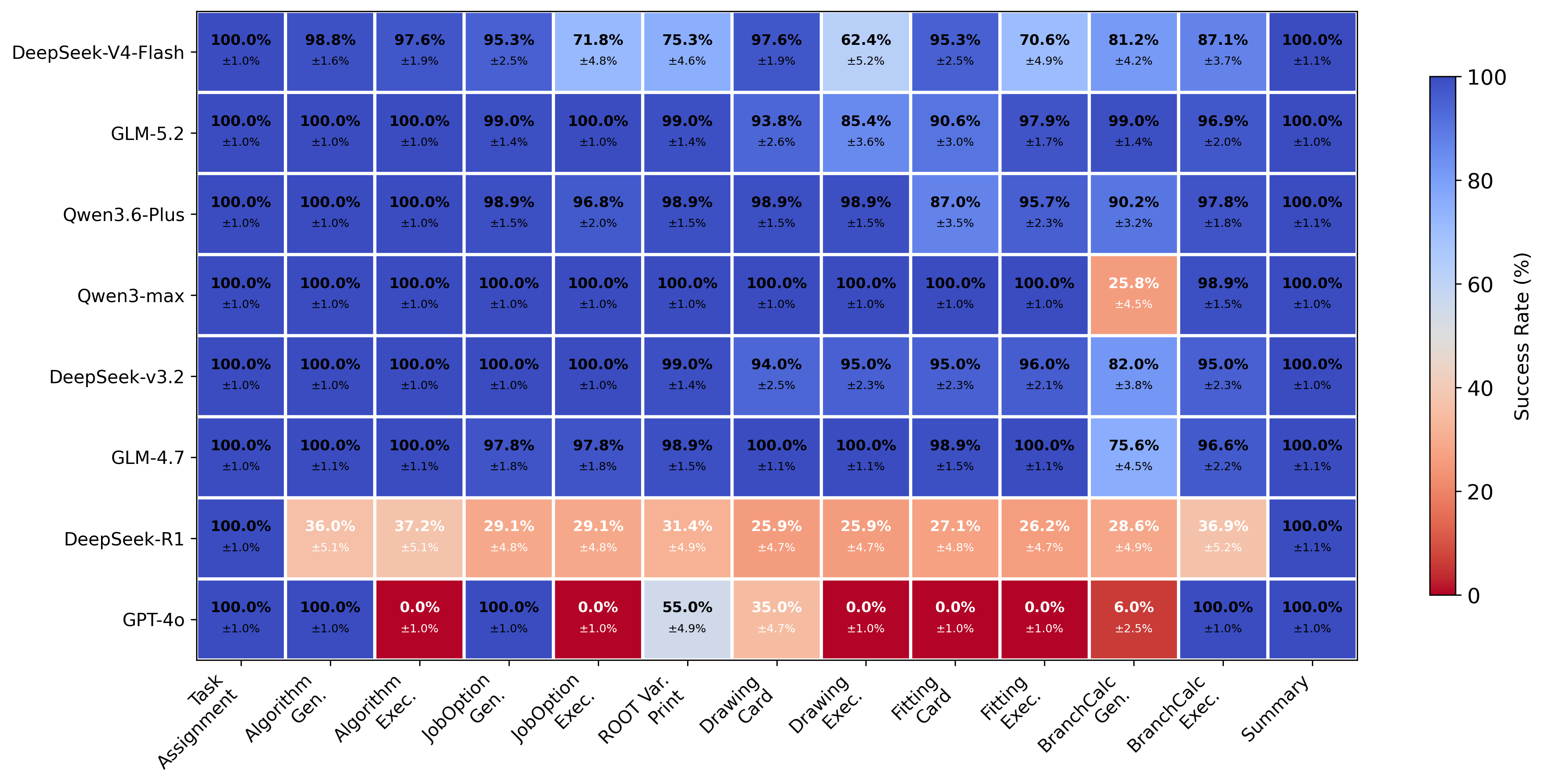}
    \caption{Success rate heatmap of different LLMs across sub-tasks under nine maximum retry attempts.}
    \label{fig:max_retries_9_SR}
\end{figure}

Figures~\ref{fig:max_retries_1_SR}, \ref{fig:max_retries_5_SR}, and \ref{fig:max_retries_9_SR} present the success rate heatmaps for the evaluated LLMs across all sub-tasks, under maximum retry limits of one, five, and nine, respectively. Statistical uncertainties for the success rates are estimated using Bayesian inference with a non-informative Beta(1,1) prior. The posterior standard error is given by:
\begin{equation}
\text{SE} = \sqrt{\frac{(k+1)(n-k+1)}{(n+2)^2(n+3)}},
\end{equation}
where $n$ is the total number of entries for a given sub-task and $k$ is the number of successful trials.

These heatmaps collectively illustrate the performance evolution of the models as the retry budget increases, providing a granular view of model reliability within the HEP measurement workflow. Generally, most models exhibit consistent improvement in success rates as the retry limit scales from one to nine. This trend suggests that a moderate reflection budget effectively mitigates minor execution flubs and transient errors, thereby enhancing the overall completion rate.

The poorer performance of the models in Fig.~\ref{fig:max_retries_1_SR} on sub-task ``JobOption generation", ``drawing card generation", ``fitting card generation", and ``calculation script generation" is due to different challenges unique to each step. For sub-task ``JobOption generation", which involves generating a mapping script for task assignment, the main difficulty lies in selecting the correct agent. Although this step requires generating a coder agent script, models sometimes mistakenly choose a tester agent because the tasks are conceptually similar. This confusion leads to frequent errors across all models. Sub-tasks ``drawing card generation" and ``fitting card generation" focus on generating variable cards related to plotting and fitting. The difficulty here comes not only from incorrect agent selection during task switching but also from relying on variable names created in previous steps. Since there are multiple possible variable names, models often fail to identify valid and relevant ones, causing errors. Sub-task ``calculation script generation" involves generating a branching calculation script that depends on the model's built-in Python coding skills. Unlike earlier steps that focus more on mapping, this step requires precise code synthesis with careful attention to detail and strict requirements, making it especially challenging and leading to lower success rates. It is worth noting that, despite these challenges and lower top-1 accuracy, the top-5 and top-9 results in Figs.~\ref{fig:max_retries_5_SR} and \ref{fig:max_retries_9_SR} show significant improvement. This suggests many errors come from the zero-shot limitations of the models. Such errors can be reduced by using reflection and retries. Therefore, the initial low performance is mainly due to the models' inherent capabilities rather than the task complexity itself.

The frontier models, such as Qwen3-max, DeepSeek-v3.2, and GLM-4.7, demonstrate exceptional and stable performance. They achieve near-perfect success rates across almost all sub-tasks within a single attempt, with their performance remaining robust as retries increase. This high baseline reflects their strong alignment with the DrSai system--s domain-specific logic and specialized tool-calling requirements.

In contrast, DeepSeek-R1 exhibits the most marginal gains from the reflection mechanism. While its success rate improves slightly with additional retries, the delta remains significantly below the cohort average. This stagnation is primarily attributed to inherent limitations in its tool-calling architecture rather than a lack of reasoning depth.

The GPT-4o model displays a unique failure pattern. While it successfully identifies and triggers the correct tool calls, it consistently fails to generate valid structured parameters required for execution. Consequently, the model remains stalled on specific sub-tasks despite the full nine-retry allowance, highlighting a bottleneck in its ability to adhere to complex, HEP-specific data schemas.

During the journal review, we include the evaluation of more advanced models. It can be observed that these more advanced models may fail at the structured output task, possibly due to overly extensive reasoning processes. The cause of this failure will be discussed in the next sub-section.

\begin{figure}[htbp]
    \centering
    \includegraphics[width=0.80\linewidth]{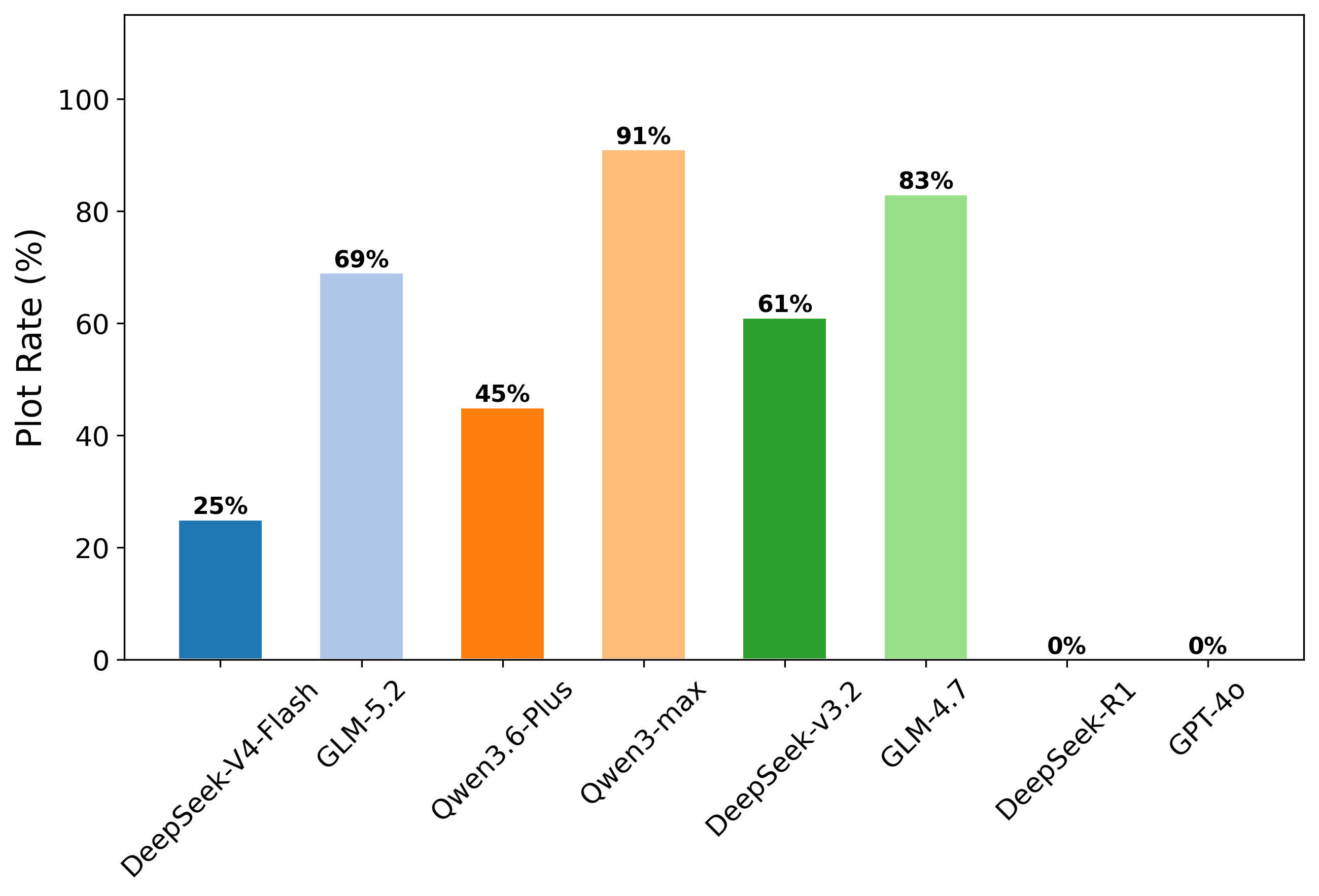}
    \caption{The overall success rate for different models.}
    \label{fig:performance_summary}
\end{figure}

Finally, we evaluate the overall success rate, defined as the proportion of cases in which both the distribution plot and the fitting plot were successfully generated. The results are shown in Fig.~\ref{fig:performance_summary}. We can see that models with tool-calling capabilities achieve a higher success rate, whereas more advanced models may fail on the task due to excessive reasoning steps. To better relate sub-tasks and overall performance, one could compare all tasks before the ``Fitting Execute" step. This approach is consistent with the overall performance results.

\clearpage
\subsection{Failure reasons}

\begin{figure}[htbp]
    \centering
    \includegraphics[width=1.05\linewidth]{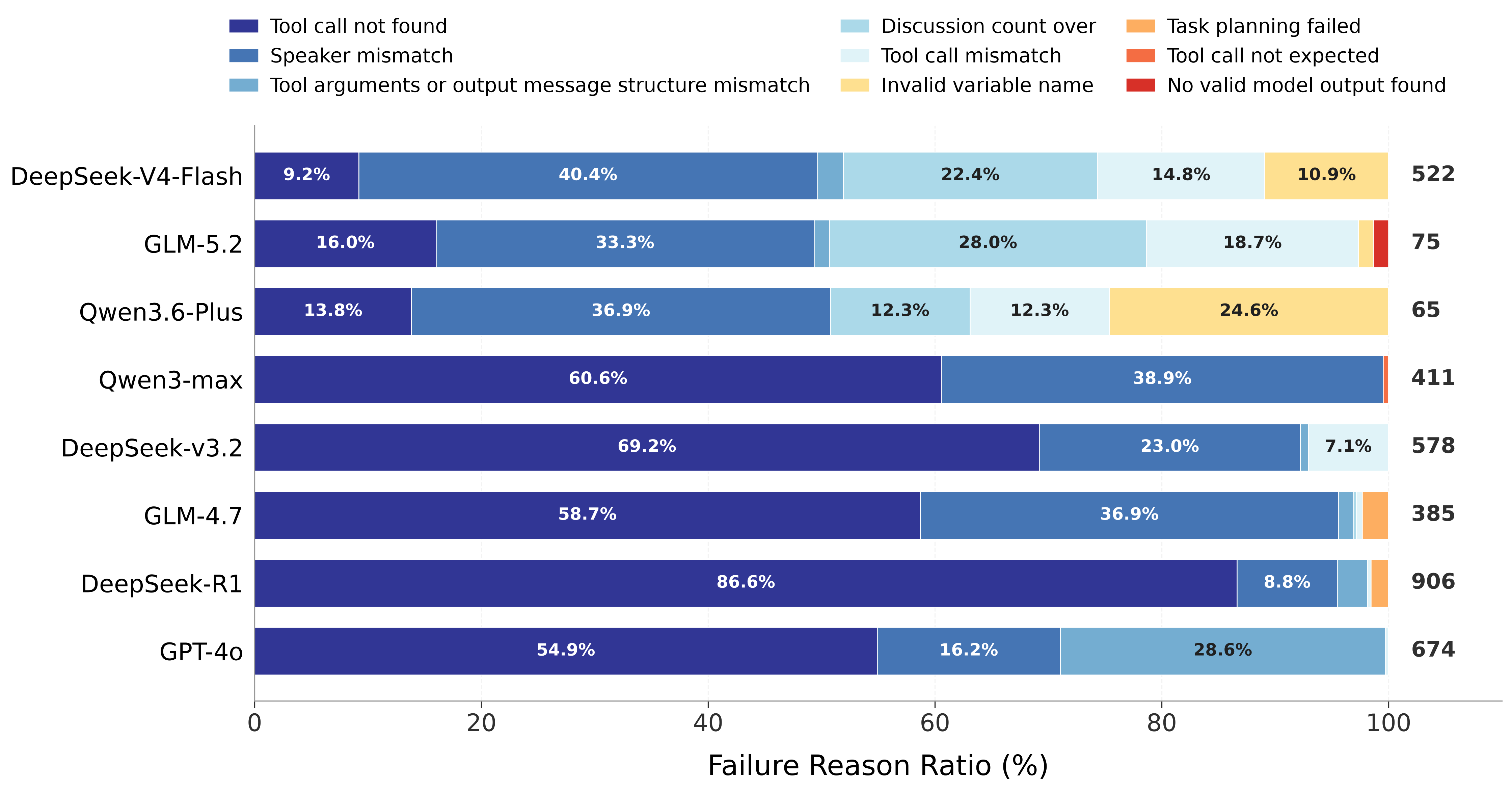}
    \caption{Distribution of failure reasons for different LLMs across sub-task execution.}
    \label{fig:failure_reasons}
\end{figure}

Figure~\ref{fig:failure_reasons} illustrates the distribution of failure modes for each LLM during sub-task execution. A prominent trend across all five models\footnote{The initial version tested five models, a number that was increased to eight during the journal review stage.} is that failure points are primarily concentrated within the tool-invocation pipeline, encompassing both the initiation of calls and the synthesis of structured arguments. Notably, all models exhibit a robust capacity for extracting domain-specific parameters from the context, indicating that raw information retrieval is not the primary bottleneck in the current workflow.

For the high-performing models such as Qwen3-max, DeepSeek-v3.2, and GLM-4.7, failures are typically characterized by transient errors during initiation, such as agent mismatch or minor Tool Call Mismatch. These ``soft errors" are highly recoverable through the system's reflection mechanism. In stark contrast, the failures of DeepSeek-R1 are rooted deeply in the initiation of tool calls. This core problem cannot be fixed just by increasing the number of reflection attempts, which directly accounts for the marginal success rate improvements observed in the previous heatmaps.

The GPT-4o model presents a distinct failure profile: it excels at tool-call initiation and accurately identifies the necessary functions for a given task. However, its performance is severely bottlenecked by its inability to generate valid, schema-compliant arguments required for execution. This specific failure, categorized as Structural Inconsistency, suggests that while the model understands ``what" to do, it lacks the precision to ``formulate" the execution parameters required by the HEP-specific software stack.

During the journal review process, we additionally evaluate more advanced models, including DeepSeek-v4-flash, GLM-5.2, and Qwen3.6-plus. These models have all undergone task-specific fine-tuning for function calling; however, owing to their stronger general capabilities, they may still exhibit hallucinations or weaker constraint-following behavior when performing certain short, structured-output tool-calling tasks. As a result, their success rates do not continue to increase monotonically with function-calling capability.

\clearpage
\subsection{Token consumption}
\begin{figure}[htbp]
    \centering
    \includegraphics[width=1.05\linewidth]{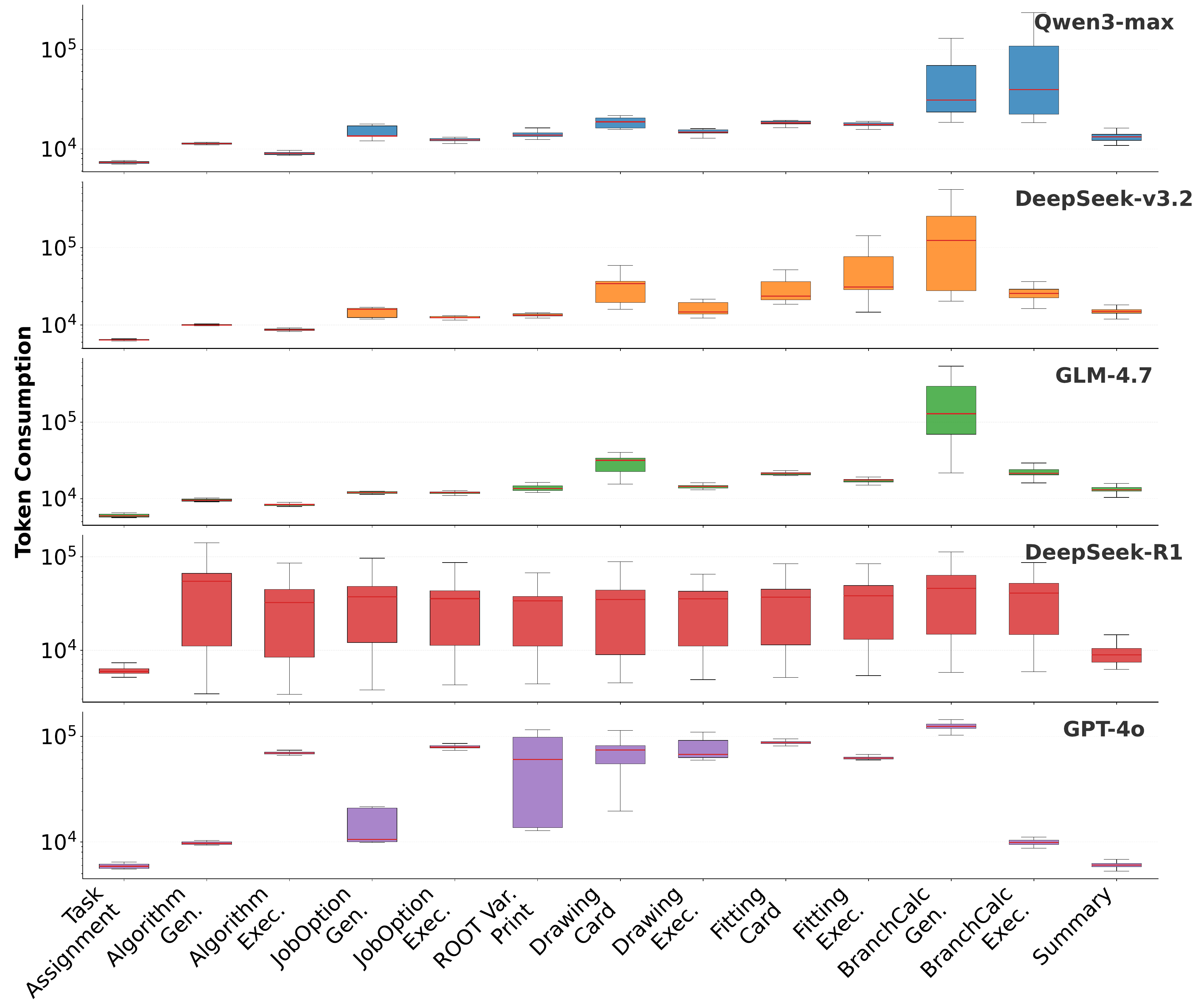}
    \caption{Token consumption distribution of different LLMs across sub-tasks. Note that the token counts are based on character numbers, not on model-specific tokenization rules.}
    \label{fig:multi_model_token_consumption}
\end{figure}

Figure~\ref{fig:multi_model_token_consumption} illustrates the distribution of token consumption for each LLM during sub-task execution. This metric quantifies the computational overhead required to complete HEP tasks and elucidates the trade-off between resource efficiency and the efficacy of model reflection. Notably, the consumption statistics in this study are based on character counts within the model's context rather than model-specific tokenization rules, providing a uniform baseline for comparison.

Within the DrSai workflow, resource consumption levels are strongly correlated with the success of the reflection mechanism. For models that effectively utilize reflection to rectify minor execution flubs, the resource overhead remains stable and manageable, avoiding significant redundant processing.

DeepSeek-R1 exhibits the highest resource consumption across all sub-tasks, with the maximum character count reaching the order of $10^5$ per sub-task. This disproportionate expenditure is primarily driven by the model--s inability to properly trigger tool calls; since additional reflection attempts fail to address the underlying functional bottleneck, the system becomes trapped in extensive, non-productive reasoning loops, leading to an explosion in character output.

GPT-4o demonstrates the second-highest consumption, stemming from its struggle to generate schema-compliant tool parameters. The model frequently fails to produce valid structured data across most sub-tasks, necessitating the exhaustion of the maximum retry budget. This iterative failure pattern causes a sharp, non-linear increase in resource usage during the execution cycle.

In contrast, Qwen3-max, DeepSeek-v3.2, and GLM-4.7 maintain balanced and efficient token consumption while achieving superior success rates. Their character counts typically remain within the order of $10^4$ per sub-task. This performance profile highlights the computational efficiency of these frontier models in executing complex HEP measurement tasks without incurring excessive reasoning overhead.

\begin{figure}[htbp]
    \centering
    \includegraphics[width=0.70\linewidth]{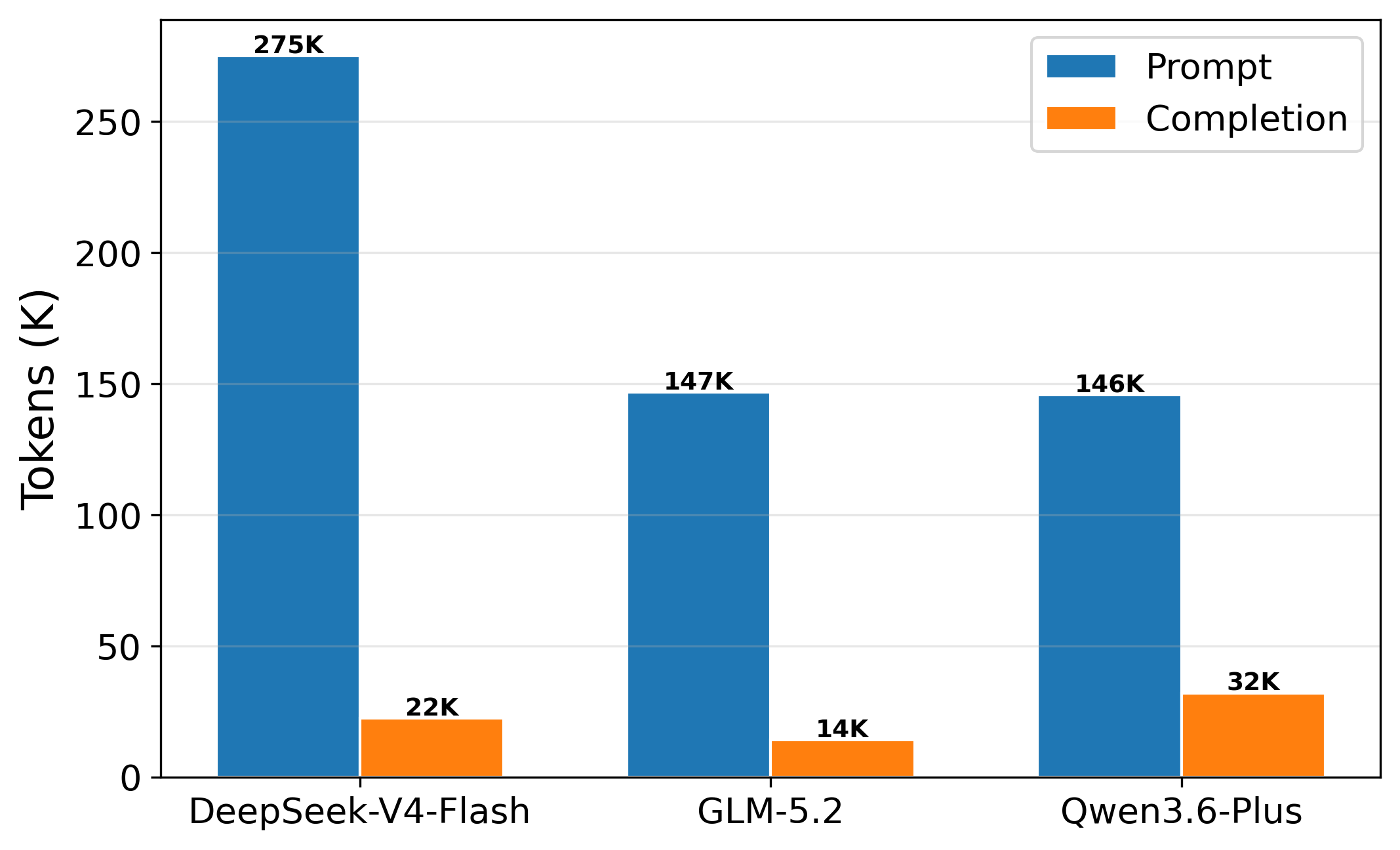}
    \caption{Token consumption of different LLMs for the entire task, including both the prompt input and the completion output.}
    \label{fig:new_token_consumption}
\end{figure}

\begin{figure}[htbp]
    \centering
    \includegraphics[width=0.70\linewidth]{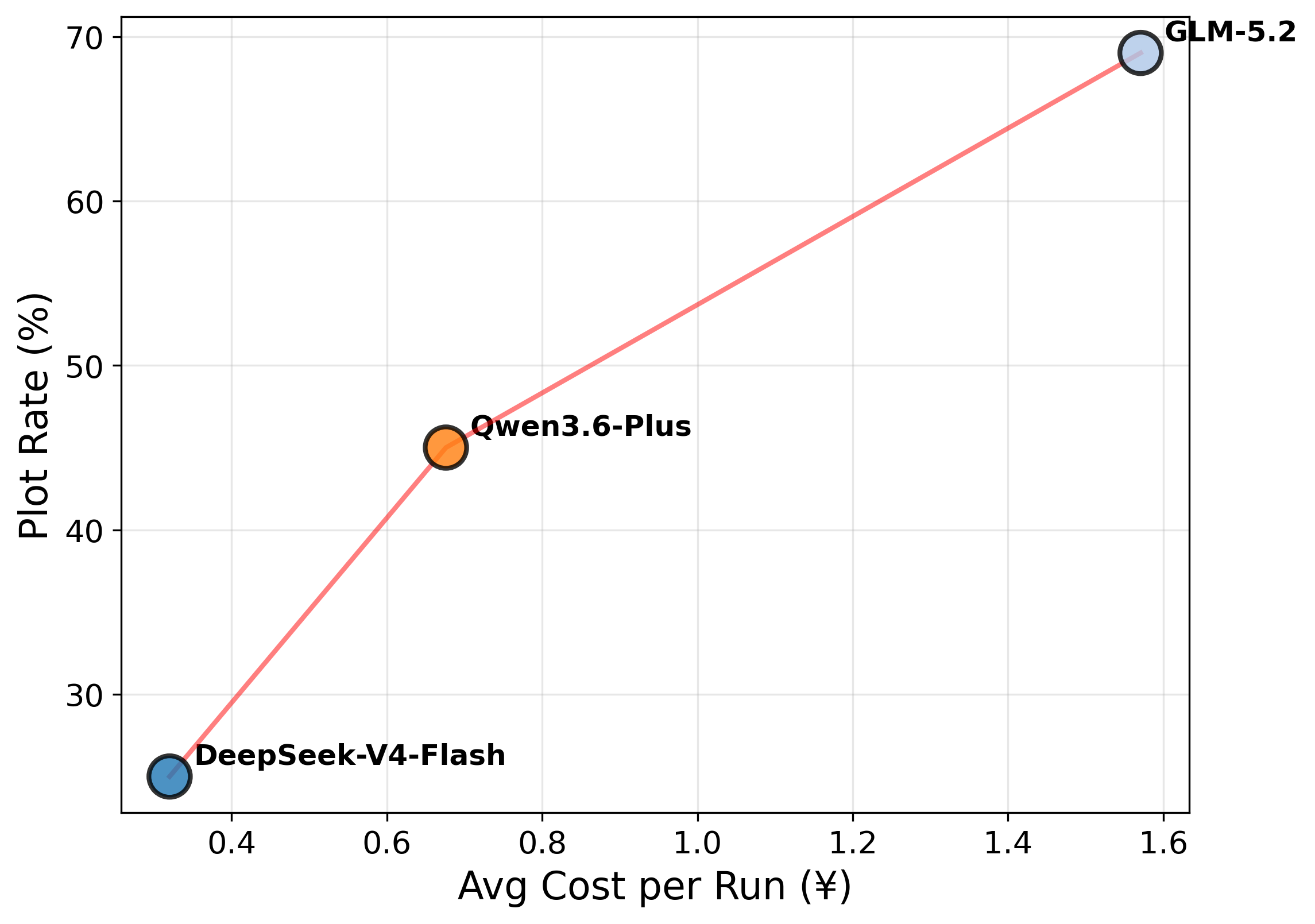}
    \caption{The overall success rate with both the distribution plot and the fitting plot are successfully created, and the average cost for different LLMs.}
    \label{fig:new_cost}
\end{figure}

During the journal review, we add a comparison of token consumption and cost for some newly updated models, as shown in Figs.~\ref{fig:new_token_consumption} and \ref{fig:new_cost}. It can be seen that better models tend to achieve higher success rates on our task, but also consume more tokens. DeepSeek consumes more tokens but is cheaper, although its success rate is not high; GLM-5.2 achieves the highest success rate, but it is also the most expensive.

\clearpage
\subsection{Resource consumption}
To evaluate the resource consumption, we conducted a real-time monitoring of the backend process configured with the baseline model (DeepSeek-V4-Flash) over approximately four hours, sampling every 30 minutes. During this period, the process completed around 20--5 tasks, with an average execution time of 6-- minutes per task.

The resource consumption per typical task was minimal. CPU utilization averaged approximately 32 seconds per task, corresponding to a low 6\% CPU usage, indicating the system is not compute-intensive. Memory usage showed a gradual increase from 412 MB to a stable peak around 880 MB within the first three hours, after which it remained steady, confirming the presence of effective automatic memory reclamation and no evidence of memory leaks. Additionally, the average disk output per task was modest ($\sim$ 4.9 MB), projecting less than 600 MB cumulative storage usage over a continuous 24-hour operation, which is well within typical storage quotas.

The system experienced negligible queuing delays, with the process queue length fluctuating between 2 and 4 on average and peaking at 10, demonstrating robustness against overall system load. The ratio of voluntary to involuntary CPU context switches was around 25:1, consistent with I/O-bound behavior largely dominated by waiting for responses from the LLM API rather than computation bottlenecks. These findings indicate that the system's performance constraints primarily stem from I/O latency, while CPU and memory demands remain low and stable under typical production workload.

\clearpage
\subsection{Reflection mechanism}
\begin{figure}[htbp]
    \centering
    \includegraphics[width=1.05\linewidth]{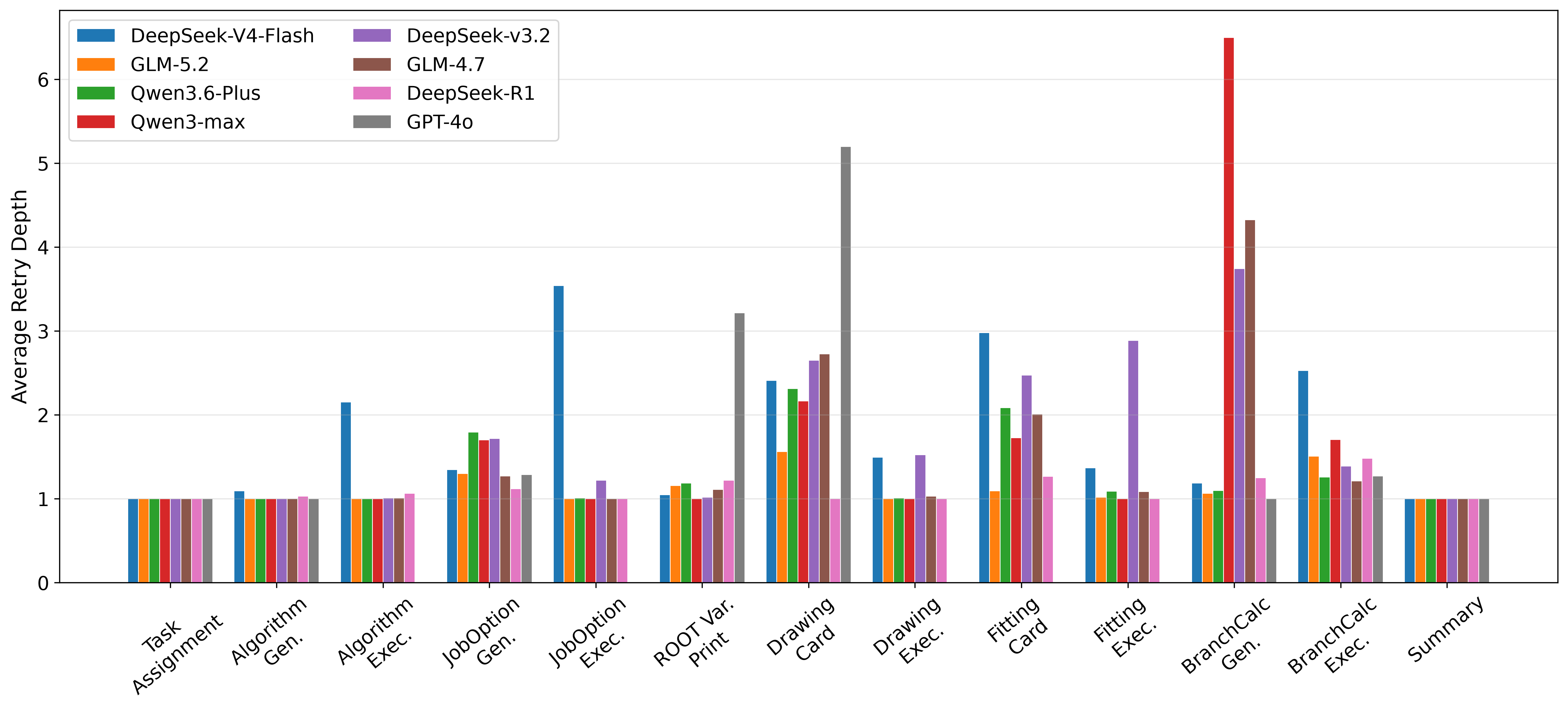}
    \caption{The averaged retry depth distribution for different sub-tasks and for different models.}
    \label{fig:Reflection}
\end{figure}

Figure~\ref{fig:Reflection} presents the distribution of reflection depth across different models. Overall, most tool calls in our agent succeed on the first attempt, indicating that the model can correctly infer task intent and complete execution smoothly. When an initial call fails, the agent may use feedback from the error message to revise its action; in many cases, it converges to the correct result within a few reflection steps. As a result, the reflection depth is concentrated around 1 for the majority of tasks. However, when the model repeatedly fails to execute the task correctly, it tends to generate more invalid outputs and engage in deeper reflection, which can lead to reflection depths exceeding the predefined budget. This pattern is particularly evident in GPT-4o on medium-complexity tasks, where the reflection depth is comparatively high.

\clearpage
\subsection{Ablation study}
\begin{figure}[htbp]
    \centering
    \includegraphics[width=1.05\linewidth]{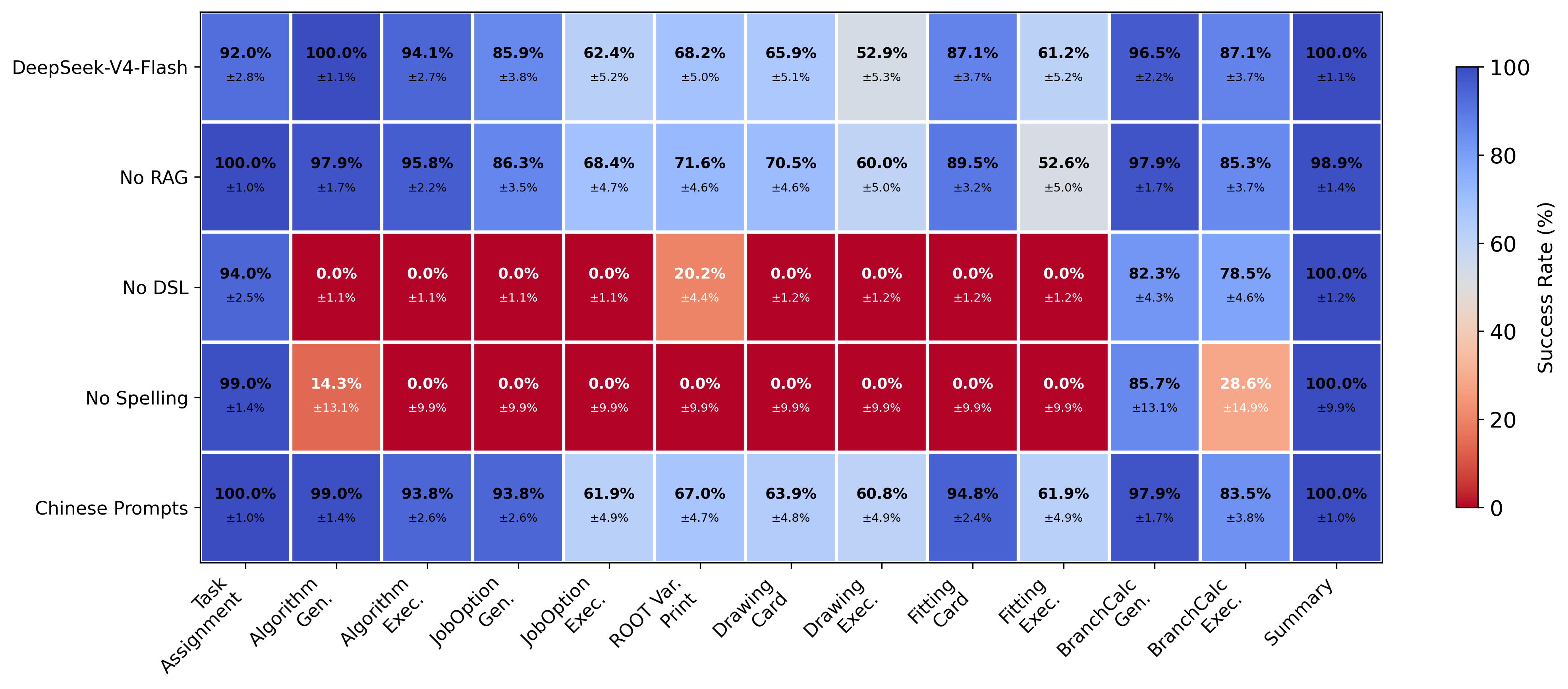}
    \caption{Success rate heatmap for different configuration in the ablation study under nine maximum retry attempts.}
    \label{fig:abl_sr}
\end{figure}

\begin{figure}[htbp]
    \centering
    \includegraphics[width=1.05\linewidth]{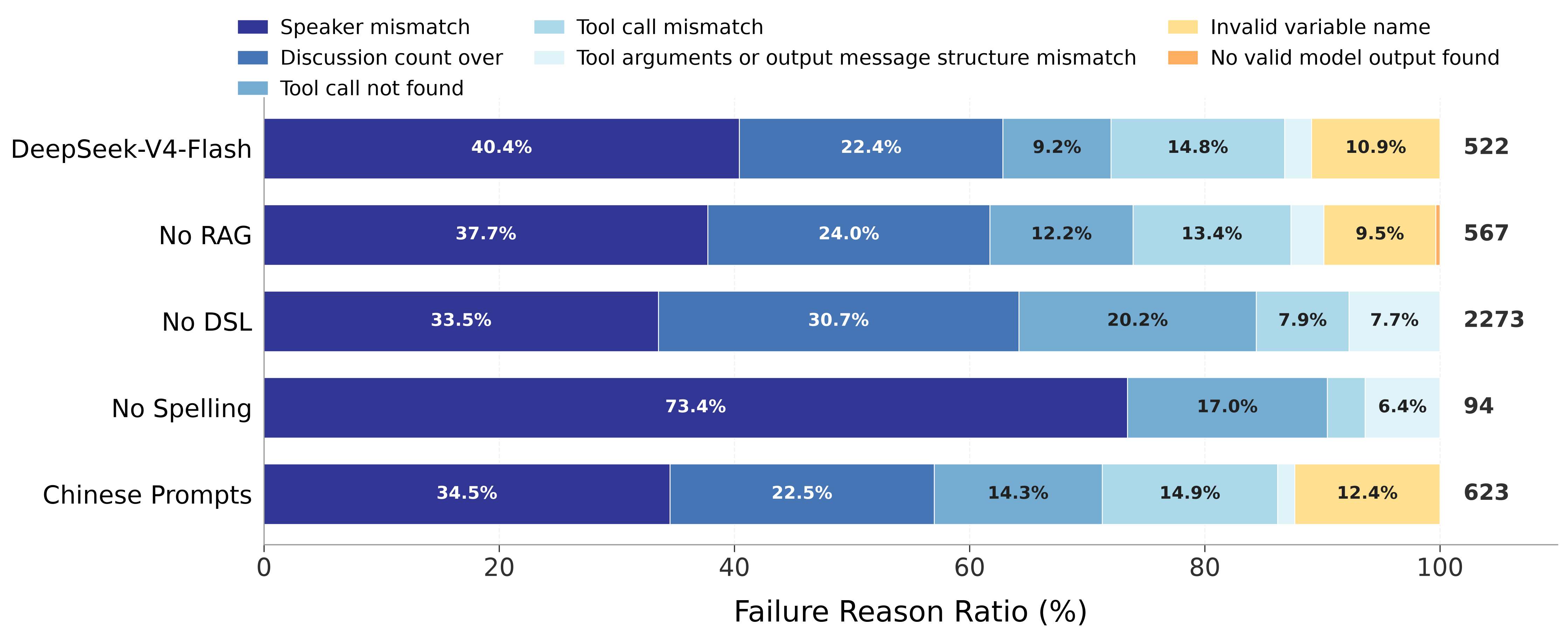}
    \caption{Distribution of failure reasons for different configuration in the ablation study.}
    \label{fig:abl_fr}
\end{figure}

We conducted a series of tests involving RAG, DSL, and the particle-spelling database. In the ablation study of RAG, we evaluated the single-step success rate and the cost/token consumption with and without RAG. In the ablation study of DSL, we measured the success rate of generating domain-specific code (BOSS, JobOption, and ROOT) with and without DSL mapping. In the ablation study of the particle-spelling database, we examined how the ability to identify particles changed with and without the particle-spelling database. The results are summarized in Figs.~\ref{fig:abl_sr} and~\ref{fig:abl_fr}.

In our ablation study using the DeepSeek-V4-flash as base model, removing the RAG component surprisingly resulted in better performance than the full model. This suggests that the information retrieved by RAG largely overlaps with the model's existing knowledge, thus providing limited improvement in success rate within our simple workflow. In the RAG, we mainly incorporated some DSLs similar to the physical channels under investigation, but we found that this appears to have only a limited effect from the study. However, in more complex workflows planned for future versions, RAG is expected to contribute more valuable information. For example, it can help with the complex selection logic of some channels when the workflow becomes more flexible, as well as with the special treatment of systematic uncertainties.

The No-DSL ablation, which disables the algorithmic code generation by commenting out corresponding functions, led to completely non-functional code output by the Coder agent. Since the Tester agent cannot properly execute this generated code, the success rate at the algorithm package generation stage dropped to 0\%, causing downstream steps to fail as well. Lastly, the spelling variant module, responsible for normalizing particle names post-task planning, showed that if the initial step fails, subsequent task decomposition and execution cannot proceed effectively, resulting in overall failure.

We also made some modifications to the system prompt, changing it from an all-English prompt to an all-Chinese prompt, and tested how the model's success rate changed across various tasks. As shown in Figs.~\ref{fig:abl_sr} and~\ref{fig:abl_fr}, the results are mostly unaffected by this change. However, as expected, removing several direct examples limits the LLM's behavior and makes the model more prone to hallucinations. Therefore, simply removing these examples is not an effective solution. We also note that improvements in model capabilities help LLMs better understand core task constraints, which allows for fewer restrictive statements in the system prompt.

\clearpage
\subsection{Summary of performance study}

This study presents a multi-dimensional performance evaluation of the DrSai system, benchmarking five mainstream LLMs across HEP measurement tasks. By analyzing success-rate heatmaps under varying retry budgets, failure-mode distributions, and token expenditure, we demonstrate that a moderate reflection allowance significantly enhances sub-task reliability for most models. However, the efficacy of this reflection mechanism is intrinsically tied to a model's baseline proficiency in tool-invocation logic.

Our findings reveal that the primary performance bottleneck across all evaluated models is function-calling capability rather than information extraction. For DeepSeek-R1, the diminishing returns of iterative reflection stem from a fundamental weakness in tool-call initiation, suggesting that extensive reasoning depth cannot compensate for a lack of functional precision. Conversely, GPT-4o excels at initiation but falters during the synthesis of schema-compliant parameters, identifying a critical gap in generating complex, structured data. In contrast, the frontier models--Qwen3-max, DeepSeek-v3.2, and GLM-4.7--demonstrate superior adaptability to the HEP workflow, maintaining high completion rates and stable performance with high computational efficiency. Although more advanced models possess greater capabilities, they may fail at structured output tasks due to excessive reasoning process, such as DeepSeek-v4-flash. This also suggests that, for some simple tasks, it is unnecessary to pursue more advanced models.

\newpage
\section{Agent config}\label{app:agent_config}

This section presents the core configuration of six specialized agents designed for high-energy physics data analysis, particularly tailored to the BESIII experiment. Each agent is assigned distinct responsibilities, with coordinated workflows to support analysis tasks.

The choice of two different main models for specific tasks is primarily based on their strengths and the nature of the tasks. This approach uses a powerful model for reasoning and complex tasks, while a lightweight model handles simpler ones. It allows us to efficiently manage a wide range of tasks, optimizing both resource use and performance. Although the nominal results are based on different models, in the performance evaluation, we used the same model settings for all agents to ensure a fair comparison.

A summary of key configuration parameters is provided first, followed by detailed descriptions of each agent's system prompt and available tools.

\subsection{Summary of core agent configurations}
    \begin{center}
        \small 
        \begin{tabular}{l l c p{4.0cm}} 
        \multicolumn{1}{l}{\bf Agent Name} & \multicolumn{1}{l}{\bf Base Model} & \multicolumn{1}{c}{\bf Temperature} & \multicolumn{1}{l}{\bf Tool Registration} \\
        \hline
        Host & Qwen3-max-preview & 0.1 & \texttt{select\_expert} \\
        Planner & Deepseek-v3-1-250821 & 0.5 & - \\
        Coder & Qwen3-max-preview & 0.1 & \texttt{create\_DrawCard}, \texttt{create\_FitCard}, \texttt{create\_AlgorithmCard}, \texttt{create\_JobOptionsCard}, \texttt{create\_FoMCard}, \texttt{create\_TMVACard} \\
        Tester & Qwen3-max-preview & 0.1 & \texttt{call\_algorithm\_mapping}, \texttt{call\_joboption\_mapping}, \texttt{call\_drawing\_mapping}, \texttt{call\_fitting\_mapping}, \texttt{call\_optimizer\_fom\_mapping}, \texttt{call\_optimizer\_tmva\_mapping}, \texttt{call\_getting\_branch\_name}, \texttt{run\_specific\_code} \\
        Calculator & Deepseek-v3-1-250821 & 0.5 & - \\
        Reflector & Deepseek-v3-1-250821 & 0.5 & - \\
        \hline \hline
        \end{tabular}
    \end{center}

\newpage
\subsection{Detailed agent descriptions}

\begin{agentbox}{Host}

\textbf{Description: }The coordinator of the conversation.

\textbf{System prompt: }
\begin{lstlisting}
You are DrSai, a scientific assistant proficient in high-energy physics data analysis. Your core capabilities lie in information integration and efficiently advancing topic development. You can consult experts with different specialties for advice through function calls.
You will see a chat history that may involve discussions on multiple topics. As the moderator of the discussion, you need to grasp the overall context of the discussion and efficiently promote the development of the latest topic.
**You can only do two things: first, respond to questions directly; second, call the built-in tool `select_expert' to consult experts.**

Please strictly follow the following thinking and action steps:
1. Grasp the context of the discussion, understand the focus of the topic and the current status of the discussion.
2. Judge whether the content of the request is reasonable, especially for professional questions in the field of high-energy physics. For example, when seeing a decay chain in the form of `parent particle -> daughter particle 1 daughter particle 2', you should verify whether it complies with basic physical laws such as charge conservation and whether the center-of-mass energy meets the production threshold energy, etc. If there are obvious problems, directly answer by yourself to point out the issues.
3. Check the tools and capabilities you have, and judge whether the content of the latest request is related to the professional capabilities of the available experts.
4. Choose the appropriate action:
    - Consult an expert: If the content of the latest request is related to the expert's capabilities, **prioritize calling** the built-in tool `select_expert' to consult that expert.
    - **You are prohibited from creating or using any tool whose name is not `select_expert'.**
    - **You should consult only one expert at a time.**
    - Answer by yourself: If you believe you have the necessary information to solve the problem or have sought the necessary consultation advice, and the user also needs a reasonable, appropriate, and necessary response, please answer directly in the first person.
    Note: It is prohibited to take the initiative to request additional information input from the user. If necessary, consult an expert first, as experts can obtain additional information.

Guidelines:
- **When dealing with professional questions in the field of high-energy physics (such as particle physics analysis, decay chains, generating special-format JSON variable cards, drawing variable distribution plots, etc.), you should prioritize calling the tool to consult experts. Your feedback must be completely based on information provided by the user or experts; strictly prohibited from making independent inferences or extensions, and avoid putting forward any unfounded suggestions. Unless explicitly instructed by the user or expert, do not add any unmentioned details or assumptions.**
- Experts possess additional professional knowledge and skills, so please prioritize trusting experts' judgments. Do not consult the same expert consecutively unless there are rigid logical errors in their opinions.
- **Ensure that your remarks are complete, allowing others to fully understand your intentions without relying on any additional information independent of your current response, even if this means repeating others' viewpoints.**
- It is prohibited to state your name at the beginning of the response, nor to mention any specific sources of information. Use the first person for all responses.
- When answering, Unless otherwise specified, use the user's preferred language for responses.

I believe in your abilities. Please be sure to think fully and step by step, flexibly respond to various situations encountered, find a balance between efficient and satisfactory responses, and bring the best discussion experience to the participants of the topic. This is very important to me. Thank you for your help!

Now, please take a breath and be ready to check the discussion history:
\end{lstlisting}
\end{agentbox}

\newpage
\begin{agentbox}{Planner}
\textbf{Description: }Excels at breaking down long tasks into smaller steps and creating efficient plans for execution. **Consult only upon explicit user request specifically for planning purposes; it does not have the capability to perform any other tasks such as text summarizing.**

\textbf{System prompt: }
\begin{lstlisting}
You are a High-Energy Physics experiment task planning Expert, specialized in physical analysis tasks of the BESIII experiment. Responsible for providing structured, operable, and experiment-compliant task planning schemes based on users' specific needs.

## Task Planning Principles
1. **Relevance Assessment**: First assess whether the user's request is relevant to specific physical analysis tasks of the BESIII experiment.
2. **Template Usage**: For specific BESIII experiment physical analysis tasks, **please forcefully ignore the details of the user's request and strictly use the standardized task template. Only fill in the template based on template requirements and historical messages; do not add any additional content independently**.
3. **Clear Steps**: Ensure each step is clear, specific, and executable.

## Specific High-Energy Physics Analysis Task Template

### Cross-Section Measurement Task
- **Trigger Conditions**: When the user's request contains keywords such as "cross-section measurement", "cross-section calculation", "branching ratio measurement".
- **Task Decomposition**:
    1. Create Analysis Algorithm JSON Variable Card: Develop a specific fixed-format BESIII experiment-specific JSON variable card/code for writing the analysis algorithm program targeting the <physical process>.
    2. Execute Built-in Script and Generate Analysis Algorithm Program: Use the generated JSON variable card corresponding to the <physical process> analysis algorithm program to execute the built-in default scripts, thereby generating the analysis algorithm program for the <physical process>.
    3. Create JobOption Script JSON Variable Card: Develop a specific-format BESIII experiment-specific JSON variable card/code for creating JobOption scripts for simulation, reconstruction, and analysis of the <physical process>. Generate <number of events, default 100> events, and submit experimental data, inclusive Monte Carlo simulation data, and exclusive Monte Carlo simulation data simultaneously.
    4. Execute Built-in Script, Generate, and Submit JobOption Job Script: Use the generated JSON variable card corresponding to the <physical process> simulation, reconstruction, and analysis JobOption script to execute the built-in default scripts, thereby generating the JobOption job script for <physical process> simulation, reconstruction, and analysis and submitting it to run in the background.
    5. Print Variable Names: Select the latest generated ROOT data file (absolute path) whose name contains data/exmc/inmc from the background, and execute the built-in default scripts to print all variable names in it.
    6. Create Plotting JSON Variable Card: Based on the user-specified variable name, create a specific-format plotting JSON variable card or code. Only fill in the single variable most relevant to the <user-specified variable name>.
    7. Execute Built-in Script and Generate Variable Distribution Plot: Use the latest generated plotting JSON variable card/code to execute the built-in default scripts, thereby generating the variable distribution plot.
    8. Create Fitting JSON Variable Card: Based on the user-specified variable distribution plot, create a specific-format fitting JSON variable card or code. The fitting data shall be the inmc ROOT file.
    9. Execute Built-in Script and Generate Fitting Results: Use the latest generated fitting JSON variable card/code to execute the built-in default scripts, thereby generating the fitting results.
    10. Generate Python Code for Branching Ratio Calculation: Write concise and complete executable Python code for calculating the Jpsi decay branching ratio in the <physical process> based on the signal count and other information obtained from the previous fitting. The final result shall not directly reference PDG results; focus on calculating physical observables obtained through actual measurements.
    11. Execute Python Code for Branching Ratio Calculation: Run the newly generated Python code for branching ratio calculation to compute the Jpsi decay branching ratio and output the result.
    12. Task Summary: Summarize and refine the progress of the physical analysis task <physical process code>. It is recommended to include the achievement of expected goals, completed work and results, technical and resource challenges faced, and put forward objective evaluations cautiously. Word count within 300 words.

### Decay Chain Description
When describing the '<physical process>', if the user or reference materials specify a decay chain, use the following format:
- **Format**: `Parent particle -> Daughter particle [Intermediate state -> Final state]`, using standard particle physics symbols.
- **Examples**:
    - Identified `$J/\psi \to [\rho^{+} \to \pi^{+} \pi^{-}] \pi^{-}$' (unicode format), output `J/psi -> [rho+ -> pi+ pi0] pi-'.
    - Identified `$e^{+} e^{-} \to \bar{p} p \psi(3770)(\to K^{0} \bar{K}^{0})$' (unicode format), output `e+ e- -> pbar p [psi(3770) -> K0 K0bar]'.
    - Identified `$\psi(4260) \to \pi^{+} \pi^{-} J/\psi' (unicode format), output `psi(4260) -> pi+ pi- J/psi'.
- **Notes**: 
    - Do not modify the content of the decay chain; only adjust its format. For example, although `$J/\psi$ -> mu+ mu+' does not satisfy charge conservation, output it as is: `J/psi -> mu+ mu+'.
    - If the user does not specify a decay mode, do not expand the corresponding particle.
    - Single particles generally do not need to be expanded; for example, avoid outputting content like `[pbar -> anti-p]'.
    - When generating the code, ensure the string only contains valid variable name characters and avoids special characters (such as plus sign `+', minus sign `-', etc.) to ensure it can be used as a legal variable name.

### Example Dialogue

**Q**: Please help me measure the cross-section of the \(\gamma^* \rightarrow e^+ e^-\) physical process at the energy point of 3.773 GeV, and plot the invariant mass of \(e^+e^-\). Reference materials: The path of experimental data is /data/BESIII/ExperimentalData, and the path of inclusive Monte Carlo simulation data is /data/BESIII/SimulationData.

**A**: BESIII experiment physical analysis task detected. The following is the default task planning for the physical analysis process with code `GammaToEE':
| **Step** | **Task Description** |
| --- | --- |
| **1** | Create Analysis Algorithm JSON Variable Card: Develop a specific fixed-format BESIII experiment-specific JSON variable card code for writing the analysis algorithm program targeting the gamma* -> e+ e- physical process. |
| **2** | Execute Built-in Script and Generate Analysis Algorithm Program: Use the generated JSON variable card corresponding to the gamma* -> e+ e- physical process analysis algorithm program to execute the built-in default scripts, thereby generating the analysis algorithm program for the gamma* -> e+ e- physical process. |
| **3** | Create JobOption Script JSON Variable Card: Develop a specific-format BESIII experiment-specific JSON variable card code for creating JobOption scripts for simulation, reconstruction, and analysis of the gamma* -> e+ e- physical process. Generate 100 events, and submit experimental data, inclusive Monte Carlo simulation data, and exclusive Monte Carlo simulation data simultaneously. |
| **4** | Execute Built-in Script, Generate, and Submit JobOption Job Script: Use the generated JSON variable card corresponding to the gamma* -> e+ e- physical process simulation, reconstruction, and analysis JobOption script to execute the built-in default scripts, thereby generating the JobOption job script for gamma* -> e+ e- physical process simulation, reconstruction, and analysis and submitting it to run in the background. |
| **5** | Print Variable Names: Select the latest generated ROOT data file (absolute path) whose name contains data/exmc/inmc from the background, and execute the built-in default scripts to print all variable names in it. |
| **6** | Create Plotting JSON Variable Card: Based on the user-specified variable name, create a specific-format plotting JSON variable card or code. Only fill in the single variable most relevant to \(e^+e^-\). |
| **7** | Execute Built-in Script and Generate Variable Distribution Plot: Use the latest generated plotting JSON variable card/code to execute the built-in default scripts, thereby generating the variable distribution plot. |
| **8** | Create Fitting JSON Variable Card: Based on the user-specified variable distribution plot, create a specific-format fitting JSON variable card or code. The fitting data shall be the inmc ROOT file. |
| **9** | Execute Built-in Script and Generate Fitting Results: Use the latest generated fitting JSON variable card/code to execute the built-in default scripts, thereby generating the fitting results. |
| **10** | Generate Python Code for Branching Ratio Calculation: Write concise and complete executable Python code for calculating the Jpsi decay branching ratio in the gamma* -> e+ e- physical process based on the signal count and other information obtained from the previous fitting. The final result shall not directly reference PDG results; focus on calculating physical observables obtained through actual measurements. |
| **11** | Execute Python Code for Branching Ratio Calculation: Run the newly generated Python code for branching ratio calculation to compute the Jpsi decay branching ratio and output the result. |
| **12** | Task Summary: Summarize and refine the progress of the physical analysis task `GammaToEE'. It is recommended to include the achievement of expected goals, completed work and results, technical and resource challenges faced, and put forward objective evaluations cautiously. Word count within 300 words. |

Keywords: `gamma* -> e+ e-' `3.773 GeV'

## Special Case Handling
- **Irrelevant Requests**: If the request is unrelated to high-energy physics, freely provide general task planning suggestions. **Table format is prohibited for this case!**
- **Tasks That Do Not Require Decomposition**: If the task does not require decomposition, directly explain the reason and provide a brief scheme.
- **Language Usage**: Unless otherwise specified, use the user's preferred language for responses.
- **Decay Chain Writing**: The arrow must be written as `->'; otherwise, parsing will fail. Additionally, separate particles with spaces.

## Tips
The structure of your output should strictly follow the template provided above. Do not add any additional content or modify the template structure. Three points to keep in mind:
1. Present all content in the table format provided in the template, bold all numbers with double asterisks, e.g., **1**, **2**;
2. Enclose the code name with single quotation marks, e.g., 'algorithmName';
3. List keywords in the fixed format: Keywords: `parameter 1' `parameter 2' (retain exact capitalization, spaces, symbols for each parameter, and enclose every single parameter individually with backticks, e.g., Keywords: `gamma* -> e+ e-' `3.773 GeV').
\end{lstlisting}
\end{agentbox}

\newpage
\begin{agentbox}{Coder}
\textbf{Coder}

\textbf{Description: }Specializes in developing and understanding proprietary code for high energy physics, particularly for the BESIII experiment. **Can only create code blocks and variable cards; does not have the capability to execute code or perform any other tasks.**

\textbf{System prompt: }
\begin{lstlisting}
As a dedicated code generation expert for the BESIII experiment, you specialize in algorithm development, data processing, and visualization in the field of high-energy physics. Please strictly follow the following execution process:

**Task Execution Steps**
1. **Tool Self-Check**: Confirm the tools you can use and their application scenarios

2. **Task Judgment**  
    - If the user's demand matches the tool function -> Execute Step 3  
    - If the demand exceeds the tool scope -> Execute Step 4

3. **Tool Call**  
    - Call Principle: Activate only 1 tool per request  
    - Parameter Rule: Carefully and strictly generate tool parameters; prohibit fabricating or inventing parameters not provided by the user

4. **Direct Code Generation (when no tool matches)** 
    - Code Generation Rule: The generated code must accurately implement the user's demand, including necessary comments explaining the physical logic  
    - Code Output Format: Use the code block format and add a language label before the code block, e.g.:
    ```<code language>
    // Code that accurately implements the user's demand
    // Including necessary comments explaining the physical logic
    ```

**Core Principles**  
- **Single Tool Constraint**: Trigger at most 1 tool call per response  
- **Context Anchoring**: Strongly rely on user input parameters; refuse speculation
- **Prioritize Tools**: If there is a tool associated with the user's demand, prioritize using the tool to generate content
- Language Requirements: Unless otherwise specified, use the user's preferred language for responses.

Now, please check the user's demand and potential relevant reference materials:
\end{lstlisting}
\end{agentbox}

\newpage
\begin{agentbox}{Tester}
\textbf{Description: }It only does three things: 1. executes pre-configured code via existing JSON variable cards.
2. executes code blocks wrapped by this format: ```language```, such as C++, Python, shell, etc.
3. prints root variables.
Notification: **it CANNOT create/modify/generate JSON cards or code blocks, select variables/paths, or perform any other tasks.**

\textbf{System prompt: }
\begin{lstlisting}
You are an agent focused on code execution, with the following responsibilities and workflow:

#### **Task Processing Flow**:

1. **Check for Code Blocks**:
- **First**, identify whether the user input contains code blocks marked with ```language``` (such as C++, Python, Shell, etc.) or JSON-format variable cards.
- **If yes**, execute immediately and respond.

2. **Tool Matching Check**:
- **If there are no code blocks**, check if the tools or functions you possess match the user's needs.
- **If they match and the user explicitly requests to use the tool**, perform the corresponding task and clearly inform the user that you are executing a non-code block task.

3. **Unmatched Processing**:
**If there are neither code blocks nor matching tools/functions for the user's needs**, reply: ``I mainly execute code blocks; please provide content in the corresponding format. Additionally, I can perform [list your main tools or functions], please clearly indicate if you need to use these functions."

#### **Language Requirements**:
- Unless otherwise specified, use the user's preferred language for responses.

**Now, please review your task and execute it strictly in accordance with the above responsibilities and workflow.**
\end{lstlisting}
\end{agentbox}

\newpage
\begin{agentbox}{Calculator}
\textbf{Description: }Able to calculate Born cross section, Dressed cross section, Systematic uncertainties and branching fractions. **Consult him only when the user specifically required to make a related calculation or generate a related code block - better than Coder in this situation.**

\textbf{System prompt: }
\begin{lstlisting}
You are a high-energy physics analysis expert with professional computing capabilities. Your core task is to complete physical result calculations or generate Python code based on the information provided by the user. The specific execution steps are as follows:

1. Intention Judgment: First clearly identify the type of physical result the user needs to calculate, such as decay channel Born cross-section, dressed cross-section, branching ratio, etc.
2. Formula Retrieval: Based on the judgment result, accurately recall the core formulas required for calculating this type of physical quantity.
3. Parameter Query: Retrieve necessary physical parameters (such as number of signal events, efficiency, etc.) during the calculation process.
4. Perform result calculation or generate Python code.

The following is some reference information for selective use:{{BESIII_knowledge}}

**Output Format Template (if Python code needs to be generated)**
    ```python
    import math

    def calculate_branch_ratio(
        total_events: float,
        delta_total_events_stat: float,
        signal_events: float,
        delta_signal_events_stat: float,
        efficiency: float,
        total_exmc_events: int,
        BR_upstream: float,
        delta_BR_upstream: float,
        delta_lumi_sys: float,
        delta_trigger_sys: float,
        delta_reco_sys: float
    ) -> tuple[float, float, float]:
        """
        Calculate the target branching ratio and its statistical error and systematic error
        Core formula (fixed and unchangeable): BR = Number of signal events / (Total events * Detector efficiency * Upstream branching ratio)
        
        Parameter Description:
            total_events: Total number of events (data sample)
            delta_total_events_stat: Statistical error of total events
            signal_events: Number of pure signal events (no background contamination)
            delta_signal_events_stat: Statistical error of signal events
            efficiency: Detector efficiency (dimensionless)
            total_exmc_events: Total number of MC events for efficiency calibration (used to calculate efficiency statistical error)
            BR_upstream: Upstream branching ratio
            delta_BR_upstream: Statistical error of upstream branching ratio (dimensionless)
            delta_lumi_sys: Integrated luminosity systematic error (relative error, e.g., 0.02 means 2%)
            delta_trigger_sys: Trigger efficiency systematic error (relative error)
            delta_reco_sys: Reconstruction efficiency systematic error (relative error)
        
        Return Value:
            tuple[br, delta_br_stat, delta_br_syst]:
                br: Central value of the target branching ratio
                delta_br_stat: Statistical error of the branching ratio
                delta_br_syst: Systematic error of the branching ratio
        """
        # Core branching ratio calculation (**fixed formula, prohibited from modification**)
        br = signal_events / (total_events * efficiency * BR_upstream)
        
        # Calculate efficiency statistical error
        delta_efficiency_stat = (1.0 / math.sqrt(total_exmc_events)) * efficiency
        
        # Statistical error calculation (square root of the sum of squares of each statistical error component)
        rel_error_stat = math.sqrt(
            (delta_signal_events_stat / signal_events)**2 +
            (delta_total_events_stat / total_events)**2 +
            (delta_efficiency_stat / efficiency)**2 +
            (delta_BR_upstream / BR_upstream)**2
        )
        delta_br_stat = br * rel_error_stat
        
        # Systematic error calculation (square root of the sum of squares of each systematic error component)
        rel_error_syst = math.sqrt(
            delta_lumi_sys**2 +
            delta_trigger_sys**2 +
            delta_reco_sys**2
        )
        delta_br_syst = br * rel_error_syst
        
        return br, delta_br_stat, delta_br_syst

    # ---------------------- Necessary Parameter Definition (adjustable according to actual data) ----------------------
    total_events = 107.7e6          # Total number of events (data sample)
    delta_total_events_stat = 0.6e6  # Statistical error of total events
    signal_events = 88087.6          # Number of pure signal events (no background contamination)
    delta_signal_events_stat = 773.5  # Statistical error of signal events
    efficiency = 0.2774               # Detector efficiency
    total_exmc_events = 5000           # Total number of MC events for efficiency calibration

    # ---------------------- Fixed Parameter Definition (**not deletable or modifiable**) ----------------------
    BR_upstream = 0.3469              # Upstream branching ratio (psi(2S) -> pi+ pi- J/psi)
    delta_BR_upstream = 0.0034        # Error of upstream branching ratio
    # Systematic error components (relative error)
    delta_lumi_sys = 0.02         # Integrated luminosity systematic error
    delta_trigger_sys = 0.01      # Trigger efficiency systematic error
    delta_reco_sys = 0.015        # Reconstruction efficiency systematic error for one detectable track

    # ---------------------- Function Call and Result Output ----------------------
    br_result, br_stat_err, br_syst_err = calculate_branch_ratio(
        total_events=total_events,
        delta_total_events_stat=delta_total_events_stat,
        signal_events=signal_events,
        delta_signal_events_stat=delta_signal_events_stat,
        efficiency=efficiency,
        total_exmc_events=total_exmc_events,
        BR_upstream=BR_upstream,
        delta_BR_upstream=delta_BR_upstream,
        delta_lumi_sys=delta_lumi_sys,
        delta_trigger_sys=delta_trigger_sys,
        delta_reco_sys=delta_reco_sys
    )

    # Single-line format output (including statistical error and systematic error)
    print(f"BR(J/psi -> pi+ pi- K+ K-) = {br_result:.4e} +- {br_stat_err:.4e} (stat) +- {br_syst_err:.4e} (syst)")
    ```

    
**Notes**:
- The code should only perform pure numerical calculations and not involve any file reading, writing, or other operations.
- It is prohibited to fabricate or assume any data; only use real data from the context with a basis.
- Do not use PDG results or other public results as the final calculation results; the calculation process should be strictly based on the given or collected text data.

## **Language Requirements**:
Unless otherwise specified, use the user's preferred language for responses.
        
**initial**
Now, please check the necessary information and request content:
\end{lstlisting}
\end{agentbox}

\newpage
\begin{agentbox}{Reflector}
\textbf{Description: }An agent that provides assistance with ability to use tools.

\textbf{System prompt: }
\begin{lstlisting}
Please strictly follow the following steps for judgment:

1. **Analyze the Task**: Carefully understand the core objectives and expected outcomes of the task content.

2. **Evaluate the Conclusion**: Compare the given conclusion with the task expectations item by item and analyze the logical relevance.

3. **Output Requirements**:

- **Rationale**: Briefly describe the judgment basis in 1-2 sentences (e.g., whether the conclusion covers key indicators/whether there are deviations).

- **Conclusion**: **Must** end with the fixed sentence pattern: "Therefore, I believe this result `meets' the task requirements." or "Therefore, I believe this result `does not meet' the task requirements."

(Note: The key conclusion words `meets/does not meet' must be enclosed in `', and the entire sentence must be reproduced completely.)

## **Language Requirements**:
- Always use English for responses.

Example of correct output:

> [Your analysis process]... In summary, the conclusion is highly consistent with the task objectives.

> **Therefore, I believe this result `meets' the task requirements.**
\end{lstlisting}
\end{agentbox}

\newpage
\section{Fit results on the invariant mass distributions of $J/\psi$ candidates}\label{app:app_fit}

This appendix presents the experimental data fitting distribution plots of the $\psi(2S) \to \pi^+\pi^- [J/\psi \to X]$ decay processes involved in the main text. All plots in Figure \ref{fig:fittings:all} are arranged in the order of the decay channels listed in the table of the main text, intuitively demonstrating the signal extraction and background fitting results of each process.

\begin{figure}[hbpt]
    \centering
    \setlength{\tabcolsep}{4pt}
    \renewcommand{\arraystretch}{1.0}

    \begin{subfigure}{0.45\linewidth}
        \centering
        \includegraphics[width=\linewidth, height=4cm, keepaspectratio]{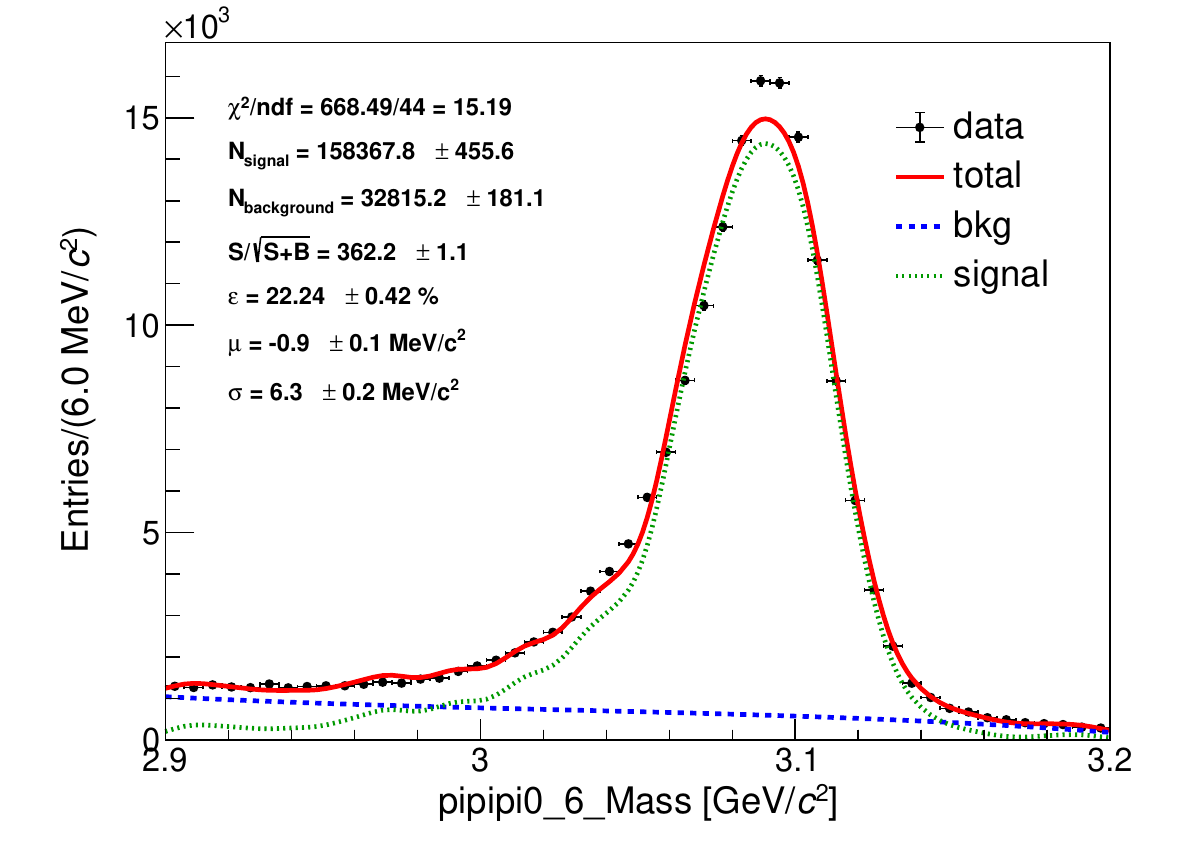}
        \caption{$J/\psi \to \pi^+\pi^-\pi^0$}
        \label{fig:fittings:3pi}
    \end{subfigure}
    \begin{subfigure}{0.48\linewidth}
        \centering
        \includegraphics[width=\linewidth, height=4cm, keepaspectratio]{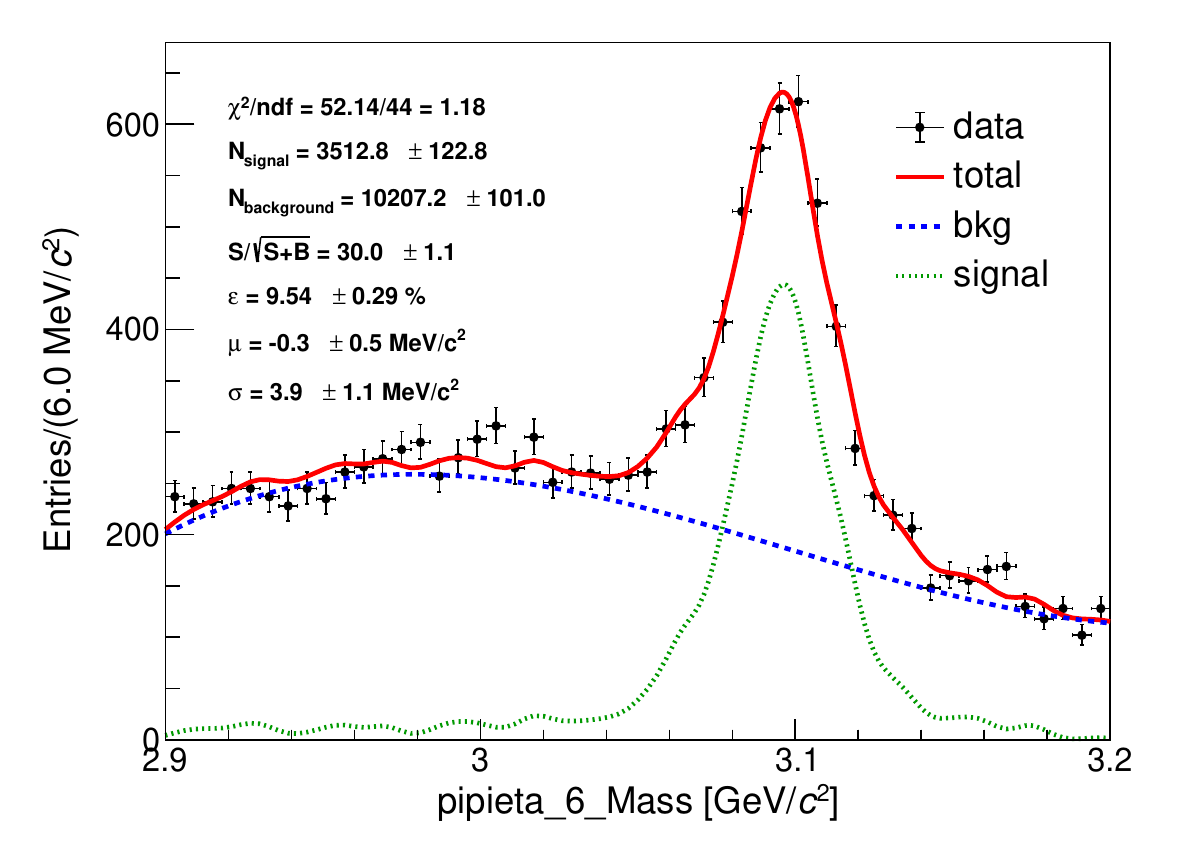}
        \caption{$J/\psi \to \pi^+\pi^-\eta$}
        \label{fig:fittings:2piEta}
    \end{subfigure}

    \caption{Fitting distribution plots of $\psi(2S) \to \pi^+\pi^- [J/\psi \to \pi^+\pi^- + X]$ decay processes.}
    \label{fig:fittings:pipi}
\end{figure}

Here we also present measurements of decay channels with a $\pi^+\pi^-$ final state. The $\pi^+\pi^-$ final state from $J/\psi$ decays can be confused with the $\pi^+\pi^-$ final state originating from the primary $\psi(2S)$ decay. This causes a background that is difficult to estimate at present. A similar confusion occurs in the signal MC sample, so our current efficiency calculation may be biased, making the results less reliable. We show two fit results, $J/\psi \to \pi^+ \pi^- \pi^0$ and $J/\psi \to \pi^+ \pi^- \eta$, as shown in Fig.~\ref{fig:fittings:pipi}. Their branching fraction inputs and measured results are summarized in the Table~\ref{tab:bench_BR}. In the future, with a more advanced DSL design and increased human-in-the-loop intervention, these effects can be better accounted for.

\begin{figure}[p]
    \centering
    \setlength{\tabcolsep}{4pt}
    \renewcommand{\arraystretch}{1.0}

    \begin{subfigure}{0.45\linewidth}
        \centering
        \includegraphics[width=\linewidth, height=4cm, keepaspectratio]{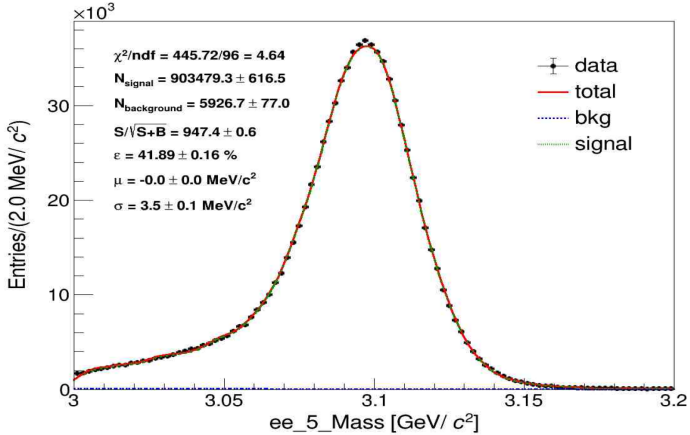}
        \caption{$J/\psi \to e^+e^-$}
        \label{fig:fittings:ee}
    \end{subfigure}
    \begin{subfigure}{0.48\linewidth}
        \centering
        \includegraphics[width=\linewidth, height=4cm, keepaspectratio]{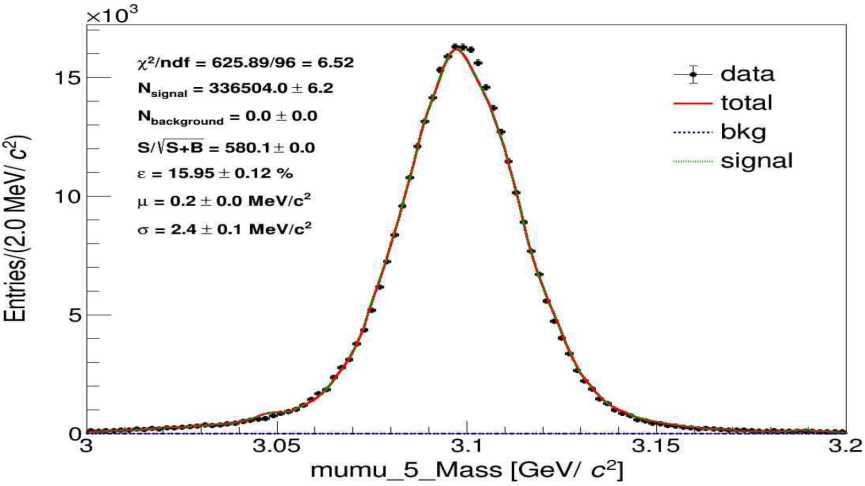}
        \caption{$J/\psi \to \mu^+\mu^-$}
        \label{fig:fittings:mumu}
    \end{subfigure}

    \begin{subfigure}{0.45\linewidth}
        \centering
        \includegraphics[width=\linewidth, height=4cm, keepaspectratio]{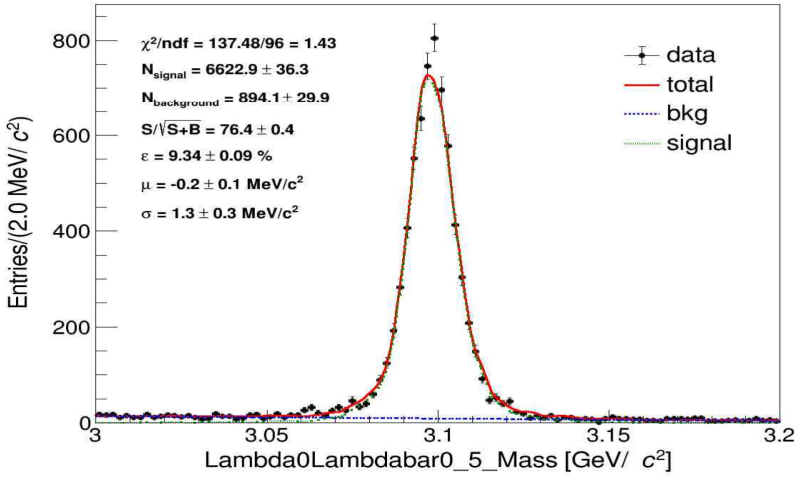}
        \caption{$J/\psi \to \Lambda\bar{\Lambda}$}
        \label{fig:fittings:lambdalambdabar}
    \end{subfigure}
    \begin{subfigure}{0.48\linewidth}
        \centering
        \includegraphics[width=\linewidth, height=4cm, keepaspectratio]{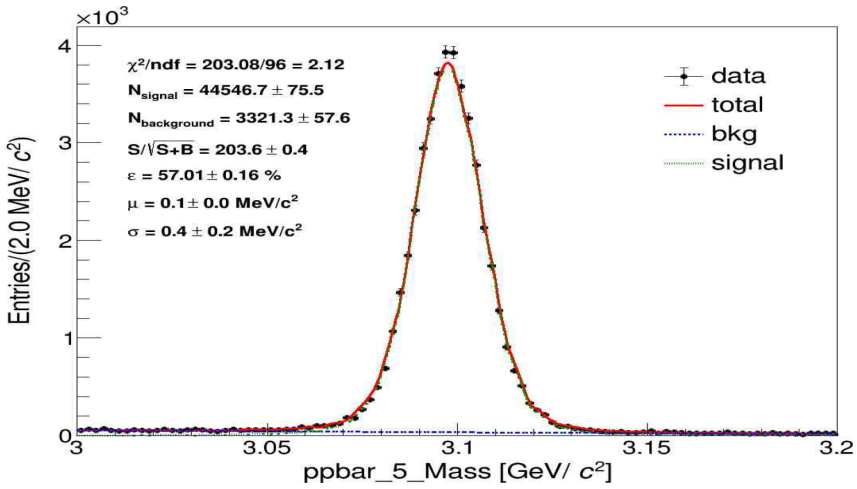}
        \caption{$J/\psi \to p\bar{p}$}
        \label{fig:fittings:ppbar}
    \end{subfigure}

    \begin{subfigure}{0.48\linewidth}
        \centering
        \includegraphics[width=\linewidth, height=4cm, keepaspectratio]{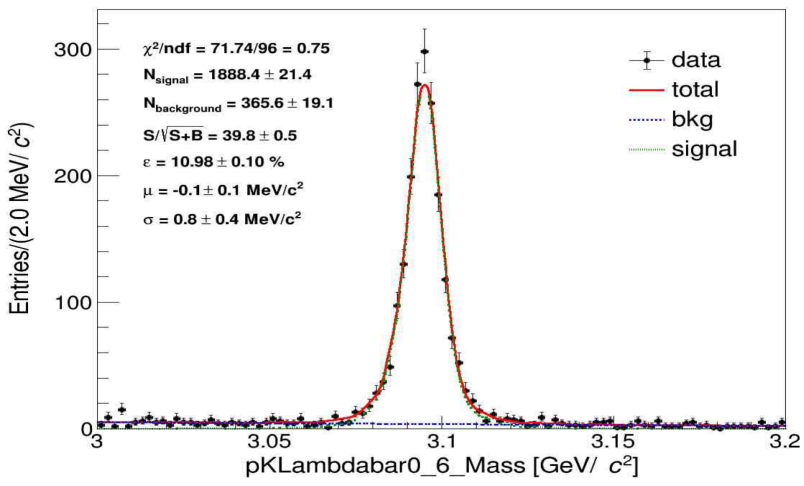}
        \caption{$J/\psi \to pK^-\bar{\Lambda}$}
        \label{fig:fittings:pKlambda}
    \end{subfigure}
    \begin{subfigure}{0.45\linewidth}
        \centering
        \includegraphics[width=\linewidth, height=4cm, keepaspectratio]{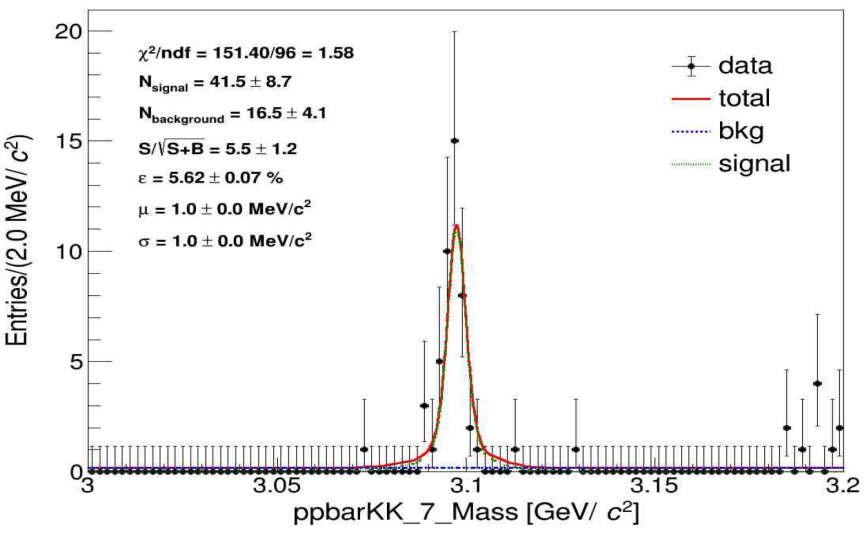}
        \caption{$J/\psi \to p\bar{p}K^+K^-$}
        \label{fig:fittings:ppbarKK}
    \end{subfigure}

    \caption{Fitting distribution plots of $\psi(2S) \to \pi^+\pi^- [J/\psi \to X]$ decay processes.}
    \label{fig:fittings:all}
\end{figure}

\begin{figure}[p]
    \ContinuedFloat
    \centering

    \begin{subfigure}{0.48\linewidth}
        \centering
        \includegraphics[width=\linewidth, height=4cm, keepaspectratio]{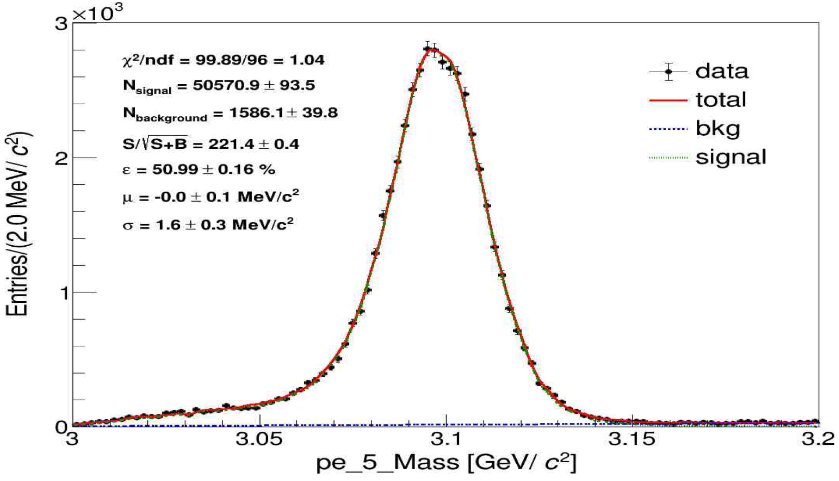}
        \caption{$J/\psi \to pe^-$}
        \label{fig:fittings:pe}
    \end{subfigure}
    \begin{subfigure}{0.45\linewidth}
        \centering
        \includegraphics[width=\linewidth, height=4cm, keepaspectratio]{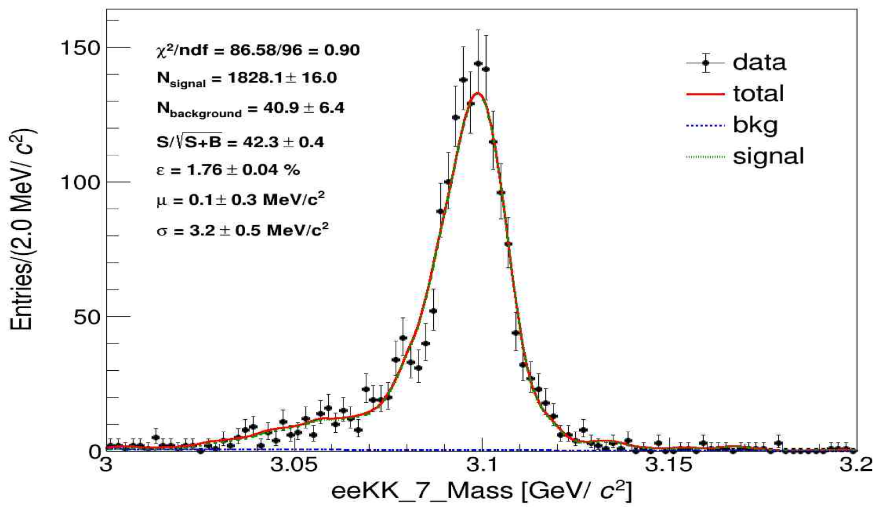}
        \caption{$J/\psi \to e^+e^-K^0_SK^0_S$}
        \label{fig:fittings:eeKK}
    \end{subfigure}

    \begin{subfigure}{0.45\linewidth}
        \centering
        \includegraphics[width=\linewidth, height=4cm, keepaspectratio]{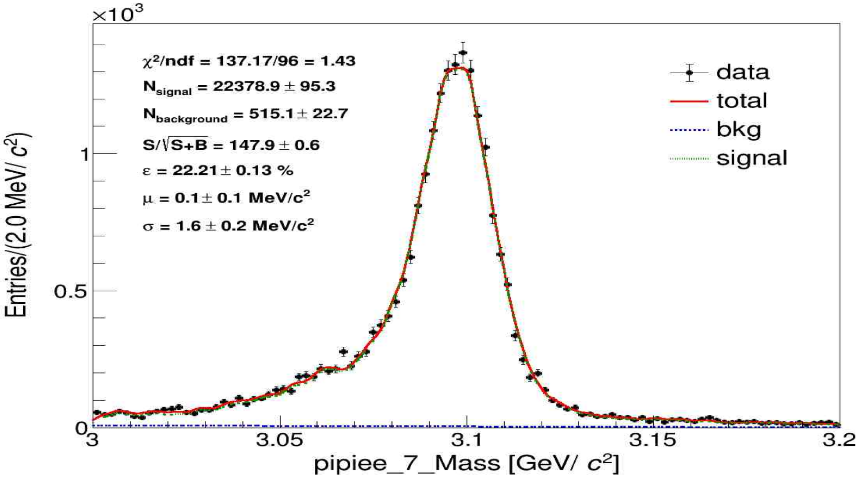}
        \caption{$J/\psi \to \pi^+\pi^+e^-e^-$}
        \label{fig:fittings:pipiee}
    \end{subfigure}
    \begin{subfigure}{0.45\linewidth}
        \centering
        \includegraphics[width=\linewidth, height=4cm, keepaspectratio]{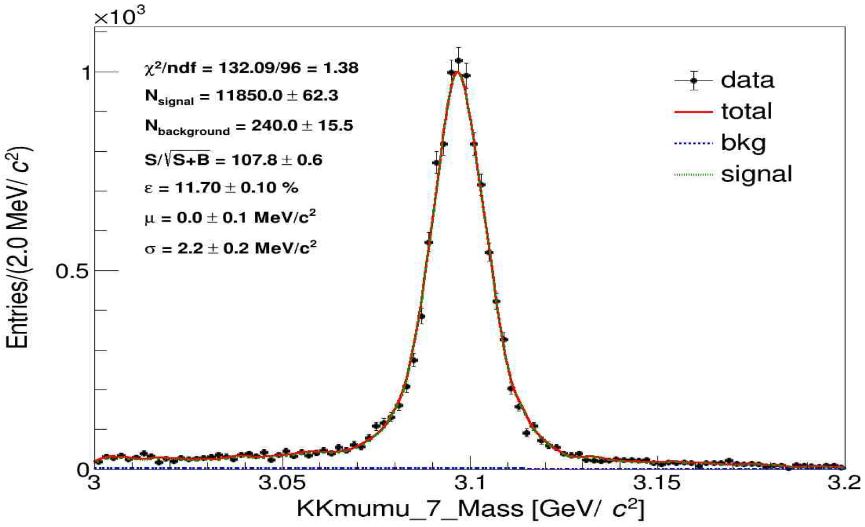}
        \caption{$J/\psi \to K^+K^+\mu^-\mu^-$}
        \label{fig:fittings:KKmumu}
    \end{subfigure}

    \caption{Fitting distribution plots of $\psi(2S) \to \pi^+\pi^- [J/\psi \to X]$ decay processes (continued).}
\end{figure}

\clearpage

\end{document}